\documentclass[twocolumn,showpacs,preprintnumbers,amsmath,amssymb,prb,superscriptaddress]{revtex4-2}
\usepackage{multirow}
\usepackage{amsfonts}
\usepackage[english]{babel}
\usepackage[T1]{fontenc}
\usepackage{times}
\usepackage{mathrsfs}
\usepackage{graphicx}
\usepackage{dcolumn}
\usepackage{bm}
\usepackage[colorlinks,bookmarks=true,citecolor=blue,linkcolor=red,urlcolor=blue]{hyperref}
\usepackage[tight, FIGTOPCAP, hang, raggedright, nooneline]{subfigure}

\begin{document}

\title{Characterization of fractional Chern insulator quasiparticles \\ in twisted homobilayer MoTe$_2$}

\author{Zhao Liu}
\email{zhaol@zju.edu.cn}
\affiliation{Zhejiang Institute of Modern Physics, Zhejiang University, Hangzhou 310058, China}
\affiliation{Zhejiang Key Laboratory of Micro-Nano Quantum Chips and Quantum Control, School of Physics, Zhejiang University, Hangzhou 310027, China}
\author{Bohao Li}
\email{bohaoli@whu.edu.cn}
\affiliation{School of Physics and Technology, Wuhan University, Wuhan 430072, China}
\author{Yuhao Shi}
\affiliation{Zhejiang Institute of Modern Physics, Zhejiang University, Hangzhou 310058, China}
\author{Fengcheng Wu}
\email{wufcheng@whu.edu.cn}
\affiliation{School of Physics and Technology, Wuhan University, Wuhan 430072, China}
\affiliation{Wuhan Institute of Quantum Technology, Wuhan 430206, China}

\date{\today}

\begin{abstract}
We provide a detailed study of Abelian quasiparticles of valley polarized fractional Chern insulators (FCIs) residing in the top valence band of twisted bilayer MoTe$_2$ (tMoTe$_2$) at hole filling $\nu_h=2/3$. We construct a tight-binding model of delocalized quasiparticles to capture the energy dispersion of a single quasiparticle. We then localize quasiparticles by short-range delta impurity potentials. Unlike the fractional quantum Hall (FQH) counterpart in the lowest Landau level (LLL), the density profile around the localized FCI quasiparticle in tMoTe$_2$ depends on the location of the impurity potential and loses the continuous rotation invariance. The FCI quasiparticle localized at moir\'e lattice center closely follows the anyon Wannier state of the tight-binding model of the mobile quasiparticle. Despite of the difference in density profiles, we find that the excess charge around the impurity potential for the $\nu_h=2/3$ FCIs in tMoTe$_2$ is still similar to that of the $\nu=2/3$ FQH state in the LLL if an effective magnetic length on the moir\'e lattice is chosen as the length unit, which allows a rough estimation of the spatial extent of the FCI quasiparticle. Far away from the impurity potential, this excess charge has the tendency to reach $e/3$, as expected for the Laughlin quasiparticle. The braiding phase of two FCI quasiparticles in tMoTe$_2$ also agrees with the theoretical prediction of fractional statistics. We characterize the interaction between two FCI quasiparticles and find a crossover from repulsive to attractive interaction as gate-to-sample distances decreases. Based on the nearly ideal quantum geometry of the top valence band of tMoTe$_2$, we propose a trial wave function for localized FCI quasiparticles, which reproduces the key feature of the density profile around a quasiparticle. 
\end{abstract}

\maketitle

\section{Introduction}
The nontrivial topology and enhanced electron-electron interactions in two-dimensional narrow bands with nonzero Chern number have given rise to various novel correlated phases of matters. As a representative example, interacting particles partially occupying a nearly flat topological band can form zero-field analogues of the celebrated fractional quantum Hall (FQH) effect~\cite{PhysRevLett.48.1559,Laughlin:1983p301} at suitable filling factors, dubbed fractional Chern insulators (FCIs) ~\cite{Sun-PhysRevLett.106.236803,Tang-PhysRevLett.106.236802,neupert-PhysRevLett.106.236804,sheng-natcommun.2.389,regnault-PhysRevX.1.021014,Parameswaran2013816,BERGHOLTZ-JModPhysB2013,LIU2024515}. A main trend in the recent development of this field is to gather inspiration from material science to look for lattice platforms that can really produce FCIs. Remarkably, FCIs were theoretically predicted to exist in van-der-Waals heterostructures with moir\'e patterns~\cite{PhysRevLett.124.106803,repellinChernBandsTwisted2020,PhysRevResearch.2.023237,zhaoTDBG,PhysRevResearch.3.L032070,PhysRevB.107.L201109,xu2025fractional}. Consistent with these theoretical predictions, experiments have reported observations of FCIs in rhombohedral graphene aligned with hexagonal boron nitride~\cite{xie2021fractional,ju2023fractional,lu2025fractional,ju2025fractional,aronson2024displacementfieldcontrolledfractionalchern,xie2025tunablefractionalcherninsulators} and twisted homobilayer MoTe$_2$ (tMoTe$_2$)~\cite{xu2023fractional,jie2023fractional,xu2023fractional2,li2023fractional,shen2024fractional,andrea2024fractional,park2025observationhightemperaturedissipationlessfractional,xu2025signaturesunconventionalsuperconductivitynear}. Further theoretical studies flourish to understand the exciting experimental results~\cite{dong2023theory,dong_anomalous_2023,PhysRevLett.133.206504,soejima2024anomaloushallcrystalsrhombohedral,PhysRevB.110.075109,DiXiao_FCI_MoTe2,PhysRevB.108.085117,PhysRevResearch.5.L032022,PhysRevB.109.045147,abouelkomsan2023band,xu2024maximally,PhysRevLett.132.096602,PhysRevB.109.205121,PhysRevB.109.205122,PhysRevB.110.205130,PhysRevB.110.245115,kwan2023moirefractionalcherninsulators,yu2024moirefractionalcherninsulators,li2025multibandexactdiagonalizationiteration,nosov2025anyonsuperconductivityplateautransitions,shi2025anyondelocalizationtransitionsdisordered,Lu2024PhysRevLett.133.186602,lu2025electromagnetic}.

While exhibiting notable differences from FQH states, FCIs carry intrinsic topological orders~\cite{TopoOrder_Wen} like their FQH cousins. The defining features of an intrinsic topological order include the quantized Hall resistance, topological ground-state degeneracy~\cite{TopoDeg} and fractionalized quasiparticle excitations called anyons~\cite{Laughlin:1983p301,Moore:1991p165,Feldman_2021}. On the one hand, the energy cost of creating these quasiparticles determines the stability of FCI ground states and is closely related to the temperature dependence of the longitudinal resistance measured in experiments. On the other hand, the direct characterization of quasiparticles, including their spatial extension, charge, statistics, and interaction, is highly relevant to the investigations of anyons using interferometers~\cite{PhysRevB.55.2331,PhysRevLett.94.166802,PhysRevLett.96.016802,PhysRevLett.96.016803,NAanyonWillet,PhysRevB.83.155440,PhysRevB.85.201302,Manfra2020,Mross2025}, anyon collider~\cite{anyoncollider2020,NAanyoncollider}, scanning tunneling microscope (STM)~\cite{PhysRevX.8.011037}, and optical imaging~\cite{Douglas-PhysRevA.84.053608}. In the context of FCIs, the energy gaps of various fractionalized excitations have been systematically studied for the FCIs at hole filling $\nu_h=2/3$ in tMoTe$_2$~\cite{goncalves2025spinlessspinfulchargeexcitations}. However, the characterization of (localized) FCI quasiparticles is mostly limited in toy tight-binding models ~\cite{PhysRevB.91.045126,PhysRevB.92.125105,Nielsen_2018,PhysRevA.98.063621,PhysRevA.98.063629,PhysRevB.99.045136,PhysRevResearch.2.013145,10.21468/SciPostPhys.12.3.095,PhysRevA.108.L061302,c6zs-m1wk}. To our knowledge, there has not been a previous attempt of a microscopic characterization of quasiparticles in moir\'e materials where FCIs were recently observed in experiments.

In this work, we present a detailed numerical characterization of the Abelian quasiparticles of $\nu_h=2/3$ FCIs in tMoTe$_2$. We adopt the continuum model of tMoTe$_2$ and focus on experimentally realistic twist angles near $4^\circ$. In the presence of interactions, the many-body calculations are mostly carried out using extensive exact diagonalization (ED) under the assumption of valley polarization and single-band projection. First of all, we find that the hole's real-space density of the FCI ground state, unlike its FQH counterpart, is strongly modulated within a moir\'e unit cell, while the total hole numbers in different unit cells remains identical. The density non-uniformness can be interpreted by the confinement of holes in specific regions of the moir\'e lattice. We then create quasiparticles in the FCI ground state and construct an effective tight-binding model to describe the dispersion of a single mobile quasiparticle. This dispersion, together with the spectrum of two mobile quasiparticles, can be used to estimate the interaction between two quasiparticles, which may predict whether bunches of two quasiparticles occurs. We find the signature of a crossover from repulsive to attractive quasiparticle interaction with the enhancement of screening in the Coulomb potential.

On the other hand, we study the FCI quasiparticles localized by delta impurity potentials. The density profile around a single localized quasiparticle varies with the location of the impurity potential, and is well described by the anyon Wannier states constructed from the tight-binding model of the mobile quasiparticle. The nearly ideal quantum  geometry~\cite{Ledwith2020Fractional,Wang2021Exact,PhysRevB.108.205144,Li2025Variational} of the top valence band of tMoTe$_2$ allows a variational construction of trial wave functions of FCI anyons. Monte Carlo simulation using these trial wave functions reproduce the key features of the density profile around a localized quasiparticle obtained by ED, thus justifying the validity of trial wave functions. We also estimate the charge of a localized quasiparticle and the braiding statistics of two localized quasiparticles in tMoTe$_2$. The results are consistent with theoretical expectations for the $\nu=2/3$ Laughlin state, i.e., fractional charge of $e/3$ and braiding phase $-2\pi/3$. However, the largest system size within the capability of ED is still insufficient for the full development of a quasiparticle. If the effective lattice magnetic length $\ell_0\equiv\sqrt{S_{\rm uc}/(2\pi)}$ ($S_{\rm uc}$ is the area of moir\'e unit cell)~\cite{PhysRevB.91.045126} and the magnetic length $\ell_B$ are used as the length units in tMoTe$_2$ and the lowest Landau level (LLL), respectively, the accumulated excess charge around the impurity potential in tMoTe$_2$ looks similar to that of the $\nu=2/3$ FQH state in the LLL, but shows slower damping with increasing distance. We estimate the quasiparticle radius of the $\nu=2/3$ FQH state in the LLL as $4\sim 6\ell_B$ by analyzing the density profile and the braiding phase. This suggests an estimation of the spatial extent of a localized $\nu_h=2/3$ FCI quasiparticle in tMoTe$_2$ as $6\ell_0$, which is $120 \ \mathring{\mathrm A}$ at twist angle $3.7^\circ$. This result could be further increased once band mixing is considered.

The remaining part of this paper is organized as follows. In Sec.~\ref{sec:continuum}, we characterize the quasiparticle of the $\nu=2/3$ FQH state in the LLL for later comparison with the FCI case. In Sec.~\ref{sec::tmote2}, we introduce the continuum model of tMoTe$_2$, and show the density of the $\nu_h=2/3$ FCI ground states. In Sec.~\ref{sec:fci}, we construct the tight-binding model of mobile quasiparticles of $\nu_h=2/3$ FCIs in tMoTe$_2$, then describe the numerical characterization of localized quasiparticles. Trial wave functions for FCI anyons and their Monte Carlo simulations are discussed in Sec.~\ref{sec::motecarlo}. Finally, we summarize our results in Sec.~\ref{sec:conclusion}. More numerical data are presented in the Appendix.

\section{FQH quasiparticles at $\nu=2/3$ in the LLL}
\label{sec:continuum}
Before studying the $\nu_h=2/3$ FCI quasiparticles in tMoTe$_2$, we first characterize quasiparticles of the conventional FQH state at $\nu=2/3$ in the LLL, for the sake of comparing quasiparticle physics in moir\'e flat bands and in the LLL. Note that we will focus on $\nu=2/3$ quasiparticles rather than quasiholes for both FQH and FCI states. We find that localized $\nu=2/3$ quasiholes have too large spatial extent so that ED cannot characterize them well in finite systems (see Appendix~\ref{app::qh}).

\subsection{Model of the LLL}
We consider $N$ interacting fermions of charge $e$ on a square torus of length $L$ penetrated by a uniform perpendicular magnetic field $B$. The number of flux quanta, $N_\phi$, through the surface of the torus is an integer satisfying $L^2=2\pi \ell_B^2 N_\phi$, where $\ell_B=\sqrt{\hbar/(eB)}$ is the magnetic length. As $N_\phi$ is also the number of single-particle states (called orbitals) per Landau level, the filling factor is defined as $\nu=N/N_\phi$. In a strong magnetic field, we can restrict all fermions in the LLL, such that the kinetic energy is simply a constant. In this case, the effective part in the many-body Hamiltonian is just the (two-body) interaction between fermions, which takes the form of
\begin{eqnarray}
H_{\rm int}=\sum_{m_1,\cdots,m_4=0}^{N_\phi-1} V_{m_1,m_2,m_3,m_4}c_{m_1}^\dagger c_{m_2}^\dagger c_{m_3} c_{m_4}.
\end{eqnarray}
Here $c_m^\dagger$ ($c_m$) creates (annihilates) a fermion in the $m$th LLL orbital, whose wave function is 
\begin{eqnarray}
\psi_m(x,y)=\left(\frac{1}{\sqrt{\pi}L\ell_B}\right)^{1/2}\sum_{n=-\infty}^{+\infty}e^{i\frac{2\pi}{L}(m+n N_\phi)y}\nonumber\\
\times e^{-\frac{1}{2\ell_B^2}\left[x-\frac{2\pi}{L}\ell_B^2(m+nN_\phi)\right]^2}
\end{eqnarray}
under the Landau gauge. The interaction matrix element is \begin{eqnarray}
V_{\{m_i\}}=\frac{1}{2}\delta_{m_1+m_2,m_3+m_4}^{{\rm mod} \ N_\phi}\sum_{s,t=-\infty}^{+\infty}\delta_{t,m_1-m_4}^{{\rm mod} \ N_\phi}\nonumber\\
\times V({\bm q})e^{-\frac{1}{2}|{\bm q}|^2\ell_B^2}e^{i\frac{2\pi s}{N_\phi}(m_1-m_3)},
\end{eqnarray}
where $\delta_{i,j}^{{\rm mod} \ N_\phi}$ is the periodic Kronecker delta function with period $N_\phi$, ${\bm q}=(q_x,q_y)=\frac{2\pi}{L}(s,t)$, and $V({\bm q})$ is the Fourier transform of the interaction potential. We consider the realistic dual-gate setup and choose the screened Coulomb potential 
\begin{eqnarray}
V({\bm q})=\frac{e^2}{4\pi\epsilon\epsilon_0}\frac{1}{L^2}\frac{2\pi\tanh(|{\bm q}|d)}{|{\bm q}|}
\end{eqnarray}
for the interaction, where $\epsilon$ is the relative dielectric constant of the material and $d$ is the distance from the top (or bottom) gate to the two-dimensional electron gas. We set $d=5\ell_B$ (weak screening) throughout our calculations in this section. The results are not sensitive to the precise value of $d$ so long as the screening is not too strong.

The ground state of the gate-screened Coulomb interaction at $\nu=2/3$ is approximately the particle-hole conjugate of the celebrated Laughlin model state, with three-fold topological degeneracy on the torus and Abelian quasiparticles. To generate $N_{\rm qp}$ quasiparticles in the ground state, we add $N_{\rm qp}$ magnetic fluxes and $N_{\rm qp}$ fermions to the system, leading to 
\begin{eqnarray}
\label{eq:NNphi}
N_\phi=\frac{1}{2}(3N-N_{\rm qp}).
\end{eqnarray}
The reason of this choice can be seen from the thin-torus configuration~\cite{PhysRevB.77.155308} of the $\nu=2/3$ Laughlin state: $110110\cdots110110$, where $1$ denotes an occupied LLL orbital and $0$ denotes an empty LLL orbital. We can create a single $111$ block by adding one occupied orbital (flux), so that the local charge density is increased by $e/3$ compared to the ground state, corresponding to a quasiparticle. The relation between $N_\phi$ and $N$ is exactly Eq.~(\ref{eq:NNphi}) if we repeat this procedure for $N_{\rm qp}$ times. 

We consider the cases with sufficiently dilute quasiparticles, such that the charge gap of the finite system remains open. In principle, we should diagonalize the full Hamiltonian $H_{\rm int}+\sum_{j=1}^{N_{\rm qp}}U_{\rm imp}({\bm w}_j)$ to get states with localized quasiparticles, where $U_{\rm imp}({\bm w})$ is an attractive impurity potential aiming to pin a quasiparticle at the specific position ${\bm w}$. However, localized quasiparticles break the translation invariance on the torus, making this direct diagonalization computationally expensive. Fortunately, as long as the the charge gap of the system remains finite, the energy spectrum of $H_{\rm int}$ always contains a low-energy subspace, corresponding to $N_{\rm qp}$ delocalized quasiparticles. The dimension of this subspace can be predicted by Haldane’s exclusion principle~\cite{Bernevig-2012PhysRevB.85.075128} or the conformal field theory~\cite{ardonne-2008-04-P04016}. Assuming that the impurity potentials only mix states within this low-energy subspace without coupling them to higher excited levels, we first diagonalize $H_{\rm int}$ to obtain the states of delocalized quasiparticles, then diagonalize the impurity potentials in the corresponding subspace to localize quasiparticles. The advantage of this two-step method is that the translation invariance is still available in the first step to reduce the many-body Hilbert space dimension. For the $\nu=2/3$ Coulomb ground state, we find an attractive $\delta$ impurity potential 
\begin{eqnarray}
U_{\rm imp}({\bm w})=-U_0\sum_{i=1}^N \delta({\bm r}_i-{\bm w}),
\end{eqnarray}
where ${\bm r}_i$ is the position of the $i$th fermion and $U_0>0$, can readily pin a quasiparticle at ${\bm w}$. The second quantization form of such an impurity potential projected to the LLL is 
\begin{eqnarray}
U_{\rm imp}({\bm w})=-U_0\sum_{m_1,m_2=0}^{N_\phi} \left[\psi^*_{m_1}({\bm w})\psi_{m_2}({\bm w}) \right] c_{m_1}^\dagger c_{m_2}.
\end{eqnarray}
Note that the value of $U_0$ cannot affect the final result of our two-step method, because only the impurity potential is involved in the second diagonalization, for which $U_0$ is just a prefactor. 

\subsection{Single quasiparticle}
\label{continuuma}

We set $N_\phi=\frac{1}{2}(3N-1)$ to generate a single quasiparticle excitation in the $\nu=2/3$ Coulomb ground state. In this case, the low-energy subspace of $H_{\rm int}$ corresponding to a delocalized quasiparticle contains $N_\phi$ states. Without loss of generality, we put the single impurity potential at the center of the torus. After diagonalizing the impurity potential in the quasiparticle subspace of dimension $N_\phi$, we obtain three nearly degenerate ground states, over which we compute the average real-space density distribution $\rho({\bm r})$ of fermions. Note that the density profile of the $\nu=2/3$ quasiparticle can be related to that of the $\nu=1/3$ quasihole by the particle-hole transformation, which is guaranteed by the particle-hole symmetry within the LLL.

\begin{figure}
\centerline{\includegraphics[width=\linewidth]{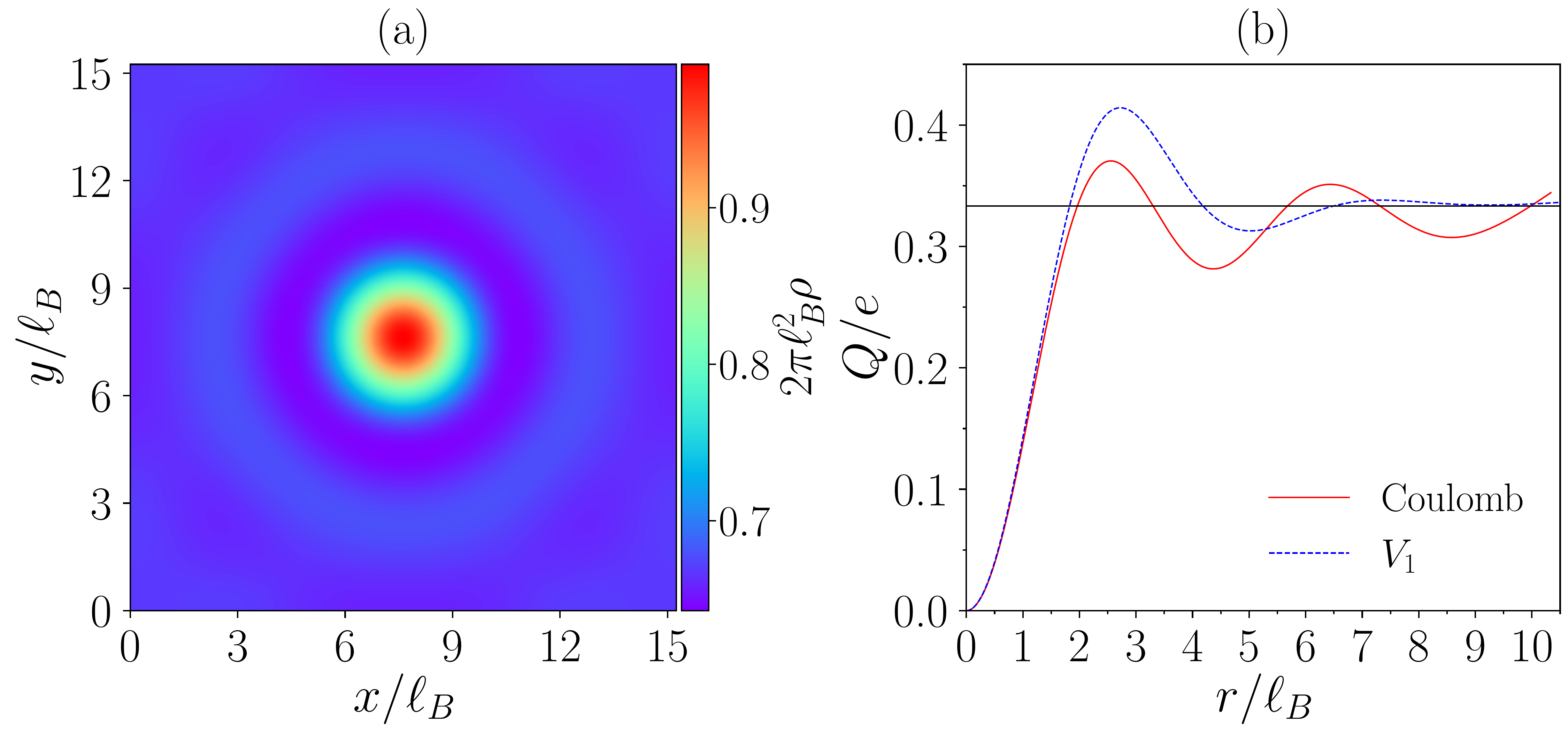}}
\caption{(a) The density distribution for a localized single-quasiparticle excitation of the $\nu=2/3$ fermionic FQH state on a square torus with $N=25,N_\phi=37$.
(b) The excess charge around the impurity potential. The horizontal reference line indicates $Q=e/3$. The dashed line in (b) is obtained using the first-order Haldane's pseudopotential. Otherwise the interaction is the screened Coulomb potential.}
\label{qh1qp}
\end{figure}

In Fig.~\ref{qh1qp}, we display the results for the largest system size, $N=25, N_\phi=37$, that we can reach using ED. There is a pronounced peak at the location of the impurity potential [Fig.~\ref{qh1qp}(a)]. To verify that the density peak corresponds to a localized quasiparticle, we compute the excess charge 
\begin{eqnarray}
Q(r)=e\int_{S_r} [\rho({\bm r}')-\rho_0({\bm r}')] d^2{\bm r}',
\label{eq::Qr}
\end{eqnarray}
where the area integral is over the disk $S_r$ of radius $r$ centered at the impurity potential, and $\rho_0({\bm r})$ is the density of the $\nu=2/3$ ground state without quasiparticles. In the FQH case, we simply have uniform $\rho_0=\nu/(2\pi\ell_B^2)$. Using the fact that $\rho({\bm r})$ is isotropic around the impurity potential, we can simplify the excess charge as $Q(r)=2\pi e\int_0^r [\rho(r')-\rho_0] r'dr'$, where $\rho(r)$ is the density distribution along the diagonal direction of the sample. One can see that $Q(r)$ oscillates around $e/3$ when $r\gtrsim 2\ell_B$ [Fig.~\ref{qh1qp}(b)]. Moreover, the oscillation amplitude decays with increasing $r$. While larger system sizes are still needed for the full convergence of excess charge, our results provide compelling evidence that a $e/3$ quasiparticle is localized by the impurity potential at the sample center, which is consistent with the Abelian $\nu=2/3$ quasiparticle.

One method to estimate the radius $R_{\rm qp}$ of the spatial extent of a quasiparticle is to utilize the second moment of the radial density $\rho(r)$~\cite{Johri-PhysRevB.89.115124,PhysRevB.91.045126,PhysRevB.99.045136}, defined as 
\begin{eqnarray}
R_{\rm qp}=\sqrt{\frac{\int_0^{r_{\max}}|\rho(r)-\rho(r_{\max})|r^3 dr}{\int_0^{r_{\max}}|\rho(r)-\rho(r_{\max})|r dr}},
\label{rqh}
\end{eqnarray}
where $r_{\max}$ is the largest distance from the torus center. A numerical calculation of Eq.~(\ref{rqh}) for the system size in Fig.~\ref{qh1qp} shows $R_{\rm qp}\approx 3.9\ell_B$. If we replace the Coulomb interaction with the short-range first-order Haldane's pseudopotential~\cite{PhysRevLett.51.605}, we get $R_{\rm qp}\approx 2.6\ell_B$. We have also checked other system sizes with $N=19$, $21$, and $23$ particles, and get almost the same $R_{\rm qp}$ from radial density for both the screened Coulomb and pseudopotential interactions. The smaller spatial extent of the quasiparticle for the pseudopotential interaction can also be seen in the excess charge [dashed lines in Fig.~\ref{qh1qp}(b)], which almost converges to $e/3$ for the same system size.

\subsection{Braiding of two quasiparticles}
\label{continuumb}
Fractional braiding statistics is another defining feature of quasiparticles. For simplicity, we generate two quasiparticles and use two impurities to pin and separate them. In this case, we again obtain three nearly degenerate ground states after diagonalizing the impurity potentials in the subspace of two delocalized quasiparticles. In order to braid these two quasiparticles, we fix the position of one impurity potential at the torus center and slowly dragging the other counter-clockwisely along a circle of radius $D$ around the torus center. The Berry phase accumulated during this process is encoded in the unitary matrix
\begin{eqnarray}
\mathcal{B}=\exp\Big\{ i\int_{0}^{2\pi}\gamma(\theta) d\theta\Big\},
\label{berry}
\end{eqnarray}
where $\theta$ is the polar angle of the mobile impurity potential, $\gamma_{nn'}(\theta)=i\langle\psi_n(\theta)|\nabla_{\theta}|\psi_{n'}(\theta)\rangle$ is the Berry connection matrix, and $\{|\psi_n(\theta)\rangle\}$ are the three ground states for the configuration of impurity potentials at a specific $\theta$. To remove the phase ambiguity in the states returned by numerical diagonalization, we impose a smooth gauge condition $\langle\psi_n(\theta)|\psi_{n'}(\theta+d\theta)\rangle=\delta_{nn'}+\mathcal{O}(d\theta^2)$, which leads to $\mathcal{B}_{nn'}=\langle\psi_n(0)|\psi_{n'}(2\pi)\rangle$. The eigenvalues of the $3\times 3$ matrix $\mathcal{B}$ are $(e^{ip_1},e^{ip_2},e^{ip_3})$, where $p_1$, $p_2$, and $p_3$ are the Berry phases. We find that their numerical values are very close to each other, so we average over them.

\begin{figure}
\centerline{\includegraphics[width=\linewidth]{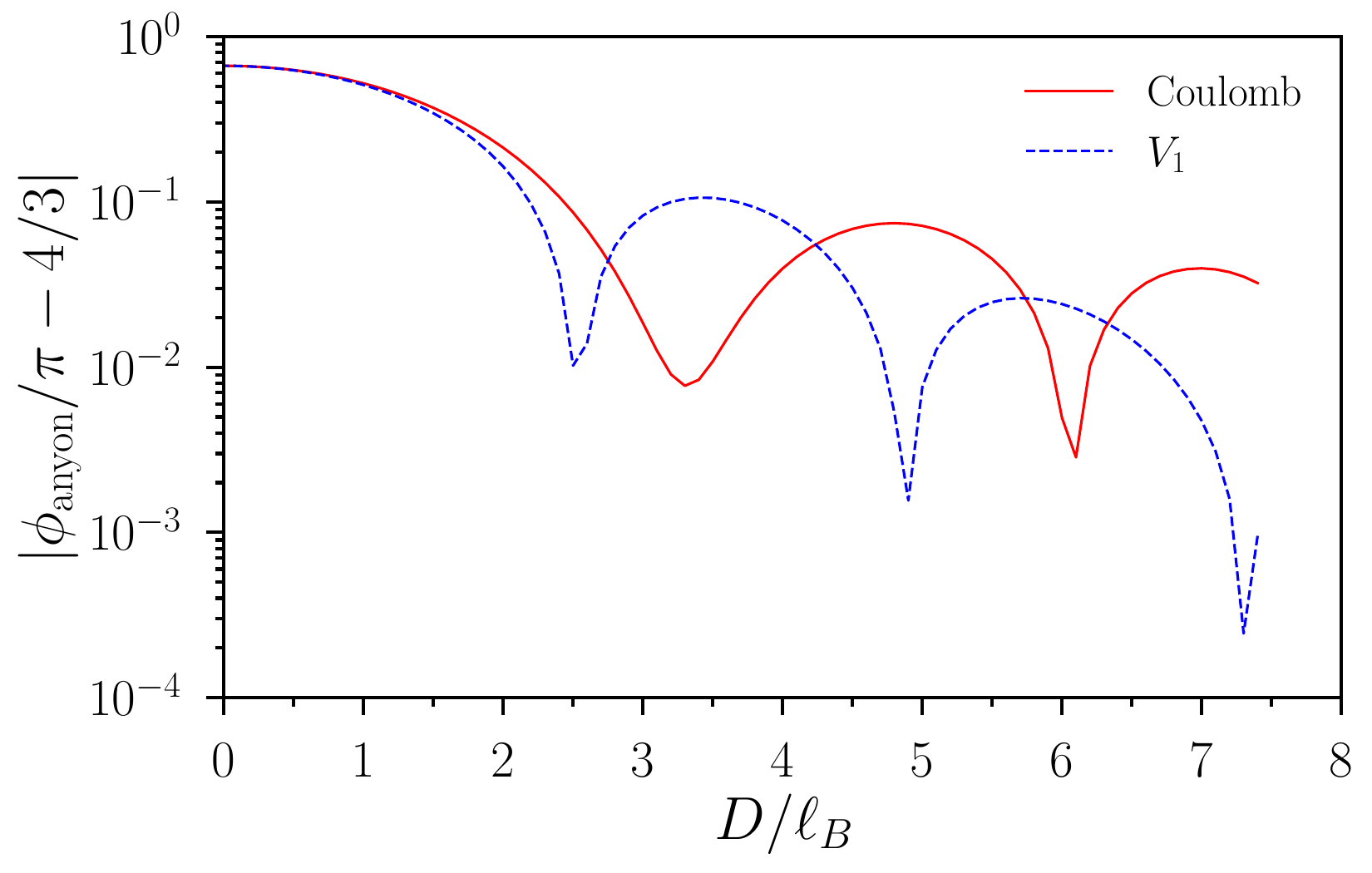}}
\caption{The fractional statistical phase versus the distance between two quasiparticles for the $\nu=2/3$ fermionic FQH state on a square torus with $N=24, N_\phi=35$. The interaction potentials are screened Coulomb and the first-order Haldane's pseudopotential for the solid and dashed lines, respectively. }
\label{qh2qp}
\end{figure}

The total Berry phase can be split into two parts: one is the Aharonov-Bohm (AB) phase caused by moving a single quasiparticle in the uniform magnetic field along the same path without the other quasiparticle enclosed; and the other comes from the fractional statistics between two quasiparticles. The AB phase $p_i^{\rm AB}$ should be $(\pi/3)(D/\ell_B)^2$, which we have numerically confirmed. Then we obtain the three fractional statistical phases as $p_i^{\rm anyon}=p_i-p_i^{\rm AB}$. Their averaged value $\phi_{\rm anyon}$ is displayed in Fig.~\ref{qh2qp} as a function of $D$ for $N=24, N_\phi=35$, which is the largest system size we can reach in the presence of two quasiparticles.
With increasing $D$, the numerically obtained anyon statistical phase demonstrates the tendency of approaching the theoretical value $4\pi/3$ (equivalently, $-2\pi/3$). The oscillation of $\phi_{\rm anyon}$ with $D$ reflects the oscillation of density profile of each quasiparticle. When $D$ is large enough, we expect that $\phi_{\rm anyon}$ exponentially converges to the theoretical value. However, as shown in Fig.~\ref{qh2qp}, the deviation from this theoretical value still oscillates around $10^{-2}\pi$ even in the largest system size for the farthest separations between two quasiparticles. Therefore, bigger samples are required to further reduce this deviation. 

The critical value of $D$ at which $\phi_{\rm anyon}$ is close enough to the theoretical value can be used as another estimation of quasiparticle size. Such a braiding-phase-based estimation is relevant to the interferometer experiment~\cite{PhysRevB.55.2331,PhysRevLett.94.166802,PhysRevLett.96.016802,PhysRevLett.96.016803,NAanyonWillet,PhysRevB.83.155440,PhysRevB.85.201302,Manfra2020,Mross2025}. We set a threshold $10^{-2}$ for the revival in the oscillations of $|\phi_{\rm anyon}/\pi-4/3|$. Using an exponential fitting for the first two revivals, we predict the critical $D\approx 11.8\ell_B$, which gives the quasiparticle radius as $R_{\rm qp}\approx 5.9\ell_B$. Compared with the Coulomb case, the pseudopotential leads to faster convergence of the fractional statistical phase (Fig.~\ref{qh2qp}), suggesting $R_{\rm qp}\approx 3.6\ell_B$. 

It is known that quantities describing the spatial extent of a quasiparticle from different aspects may give different estimations of the quasiparticle size~\cite{Johri-PhysRevB.89.115124}. As the braiding phase is sensitive to the spatial overlap of two quasiparticles, it may give a larger estimation of the quasiparticle size than using the second moment of density profile. This is consistent with our numerical results: $R_{\rm qp}$ estimated by the braiding phase is $1\sim 2\ell_B$ larger than that from the density profile. This discrepancy is within the usual range of variation among values of $R_{\rm qp}$ obtained by distinct quantities~\cite{Johri-PhysRevB.89.115124}. 

\section{Twisted bilayer MoTe$_2$}
\label{sec::tmote2}
Now we turn to FCIs in tMoTe$_2$. We will first introduce the model, then present the density of particles in the FCI ground state without quasiparticles.

\subsection{Model of tMoTe$_2$}
The moir\'e superlattice consists of two MoTe$_2$ monolayers of lattice constant $a=3.52 \ \mathring {\mathrm A}$, with a twist angle $\theta$ in between. At small twist angles, the lattice constant of the moir\'e pattern is $a_M\approx a/\theta\gg a$. Accordingly, the monolayer Brillouin zone is folded into the much smaller moir\'e Brillouin zone (MBZ). There are three high-symmetry positions in each moir\'e unit cell: $\mathcal{R}_M^M$, $\mathcal{R}_M^X$ and $\mathcal{R}_X^M$, where $M$ and $X$ represent metal and chalcogen atoms, respectively. $\mathcal{R}_\alpha^\beta$ stands for the local configuration in which $\alpha$ atom in the bottom layer is vertically aligned with the $\beta$ atom in the top layer. We choose the primitive moir\'e lattice vectors as ${\bm a}_1=(\frac{\sqrt{3}}{2},\frac{1}{2})a_M$ and ${\bm a}_2=(-\frac{\sqrt{3}}{2},\frac{1}{2})a_M$. In the reciprocal space, the $K$ and $K'$ valleys are separated by a large momentum difference at small twist angles $\theta$, such that they can be considered independently. Moreover, spin is effectively locked to valley due to the strong spin-orbit coupling near the two valleys. 

We follow a continuum description of the low-energy valence bands of tMoTe$_2$ in the $K$ and $K'$ valley~\cite{Wu_MoTe2}. Since FCIs were experimentally observed when the top moir\'e valence bands are doped by holes, it is natural to work in the hole picture. The effective single-hole continuum Hamiltonian (in the $K$ valley) takes the form of
\begin{eqnarray}
H_{0,K}({\bm r})=-
\begin{pmatrix}
    h_b({\bm r}) +\Delta_b({\bm r})& \Delta^\dagger_T({\bm r}) \\
    \Delta_T({\bm r}) & h_t({\bm r}) +\Delta_t({\bm r}) 
\end{pmatrix}
\end{eqnarray}
in the layer basis, with $t$ and $b$ denoting the top and bottom layer, respectively. The Hamiltonian $H_{0,K}({\bm r})$ differs from that in Ref.~\cite{Wu_MoTe2} by a particle-hole transformation. Here 
\begin{eqnarray}
h_l({\bm r})=-\frac{\hbar}{2m^*}(+i\nabla-{\bm K}_l)^2
\end{eqnarray}
describes the quadratic valence band edge of monolayer MoTe$_2$, where $m^*=0.6m_e$ is the effective mass and ${\bm K}_l$ is the $K$ point of layer $l$. Each layer feels a moir\'e potential 
\begin{eqnarray}
\Delta_l({\bm r})=2v\sum_{j=1,3,5}\cos({\bm G}_j\cdot{\bm r}+l \phi),
\end{eqnarray}
where ${\bm G}_j=\frac{4\pi}{\sqrt{3}a_M}\left(\cos\left(\frac{\pi}{3}(j-1)\right),\sin\left(\frac{\pi}{3}(j-1)\right)\right)$ are the shortest moir\'e reciprocal lattice vectors, and $l$ is $+1$ ($-1$)  for the bottom (top) layer. The interlayer tunneling is
\begin{eqnarray}
\Delta_T({\bm r})=w(1+e^{-i{\bm G}_2\cdot {\bm r}}+e^{-i{\bm G}_3\cdot {\bm r}}).
\end{eqnarray}
Both the intralayer moir\'e potential and the interlayer tunneling are approximated by first harmonic expansion. Throughout our work, we adopt $(v,\phi,w)=(20.8 \ {\rm meV},+107.7^\circ,-23.8 \ {\rm meV})$~\cite{DiXiao_FCI_MoTe2}, which were obtained by recent first-principles calculations for $\theta$ near $4^\circ$.  The single-particle Hamiltonian in the $K'$ valley can be determined by time reversal conjugate of $H_{0,K}({\bm r})$.

To get the moir\'e band structure, we need to further transform $H_{0,K}({\bm r})$ to the momentum space. For convenience we choose the center of the MBZ in the $K$ valley as the origin of the $k$-space. Under the plane-wave basis, the result is 
\begin{eqnarray}
H_{0,K}&=&-\sum_{\bm k}\Bigg\{\sum_{l=t,b}h_l(-{\bm k}-{\bm K}_l)c^\dagger_{K,l,{\bm k}}c_{K,l,{\bm k}}\nonumber\\
&+&v\sum_{l=t,b}\sum_{j=1,3,5}\left(e^{i l \phi}c^\dagger_{K,l,{\bm k}+{\bm G}_j}c_{K,l,{\bm k}}+{\rm h.c.}\right)\nonumber\\
&+&w\Big(c^\dagger_{K,b,{\bm k}}c_{K,t,{\bm k}}+c^\dagger_{K,b,{\bm k}+{\bm G}_2}c_{K,t,{\bm k}}\nonumber\\
&+&c^\dagger_{K,b,{\bm k}+{\bm G}_3}c_{K,t,{\bm k}}+{\rm h.c.}\Big)\Bigg\},
\label{H0h}
\end{eqnarray}
where $c^\dagger_{K,l,{\bm k}}$ creates a hole with wave vector ${\bm k}$ in valley $K$ and layer $l$. The monolayer Hamiltonian is $h_l(-{\bm k}-{\bm K}_l)=-\frac{\hbar^2|-{\bm k}-{\bm K}_l|^2}{2m^*}$. 
For each point ${\bm k}_0$ in the MBZ of the $K$ valley, we let ${\bm k}={\bm k}_0 + m_1{\bm G}_1 + m_2{\bm G}_2$ in Eq.~(\ref{H0h}). With truncations $m_1, m_2 = -M, \cdots M$ (we choose $M=7$ in our calculations), $H_{0,K}$ can be constructed as a matrix of dimension $2(2M + 1)^2$ under the plane-wave basis $|K,l,{\bm k}\rangle\equiv c^\dagger_{K,l,{\bm k}}|{\rm vac}\rangle$. The eigenvalues $E_{K,n,{\bm k}_0}$ of this matrix give the hole band energies at ${\bm k}_0$, with $n$ the band index. The band eigenvectors can be expressed as 
\begin{eqnarray}
|K,n,{\bm k}_0\rangle=\sum_{l,m_1,m_2}u_{l,m_1,m_2}^{K,n,{\bm k}_0}|K,l,{\bm k}_0 + m_1{\bm G}_1 + m_2{\bm G}_2\rangle.\nonumber\\
\end{eqnarray}
In real space, the corresponding Bloch state is a layer spinor 
\begin{eqnarray}
\psi_{K,n,{\bm k}_0}({\bm r})=\frac{1}{\sqrt{\mathcal{A}}}
\begin{pmatrix}
    \sum_{{\bm G}}e^{i({\bm k}_0+{\bm G})\cdot{\bm r}}u_{t,m_1,m_2}^{K,n,{\bm k}_0}\\
    \sum_{{\bm G}}e^{i({\bm k}_0+{\bm G})\cdot{\bm r}}u_{b,m_1,m_2}^{K,n,{\bm k}_0}
\end{pmatrix},
\end{eqnarray}
where $\mathcal{A}$ is the area of the system and ${\bm G}=m_1{\bm G}_1 + m_2{\bm G}_2$. The band structure $E_{K',n,{\bm k}_0}$ and $|K',n,{\bm k}_0\rangle$ can be obtained similarly for the $K'$ valley. For the model parameters we choose, the top valence band in each valley is isolated and relatively flat. The valence bands in different valleys carry opposite Chern number $\mathcal{C}=\pm1$. The second valence band in each valley is also isolated and topological, whose Chern number is opposite to that of the top valence band.

Given the moir\'e band structure, the whole many-body Hamiltonian of holes in tMoTe$_2$, including the band dispersion, the two-body density-density interaction of holes, and the impurity potential, can be expressed in the basis of hole bands $|\eta,n,{\bm k}\rangle$ as
\begin{eqnarray}
H&=&\sum_{{\bm k}}^{{\rm MBZ}}\sum_{\eta,n} E_{\eta,n,{\bm k}}\gamma_{\eta,n,{\bm k}}^\dagger \gamma_{\eta,n,{\bm k}}\nonumber\\
&+&\sum_{\{{\bm k}_i\}}^{{\rm MBZ}}\sum_{\eta,\eta'}\sum_{\{n_i\}}V^{(\eta\eta')}_{\{{\bm k}_i\}\{n_i\}}\gamma_{\eta, n_1,{\bm k}_1}^\dagger \gamma_{\eta', n_2,{\bm k}_2}^\dagger \gamma_{\eta', n_3,{\bm k}_3}\gamma_{\eta, n_4,{\bm k}_4}\nonumber\\
&+&\sum_{\{{\bm k}_i\}}^{{\rm MBZ}}\sum_{\eta}\sum_{\{n_i\}}U^\eta_{\{{\bm k}_i\}\{n_i\}}\gamma_{\eta, n_1,{\bm k}_1}^\dagger \gamma_{\eta, n_2,{\bm k}_2},
\label{Hh}
\end{eqnarray}
where all wave vectors are restricted in the MBZ, and $\gamma_{\eta, n,{\bm k}}^\dagger$ creates a hole with wave vector ${\bm k}$ in band $n$ of valley $\eta$, i.e., $|\eta,n,{\bm k}\rangle=\gamma_{\eta, n,{\bm k}}^\dagger|{\rm vac}\rangle$. We have assumed that the impurity potential does not flip spins. As a result of spin-valley locking in tMoTe$_2$, the impurity potential does not couple the two valleys and the number of holes in each valley is conserved. Defining 
\begin{eqnarray}
M_{\eta,n,n'}({\bm k},{\bm k}')\equiv\sum_{l,m_1,m_2}[u_{l,m_1,m_2}^{\eta,n,{\bm k}}]^*u_{l,m_1,m_2}^{\eta,n',{\bm k}'},
\end{eqnarray}
we can express the interaction matrix element as
\begin{eqnarray}
V^{(\eta\eta')}_{\{{\bm k}_i\}\{n_i\}}=\frac{1}{2}\delta'_{{\bm k}_1+{\bm k}_2,{\bm k}_3+{\bm k}_4}\sum_{\bm G}V({\bm k}_1-{\bm k}_4+{\bm G})\nonumber\\
\times M_{\eta,n_1,n_4}({\bm k}_1,{\bm k}_4-{\bm G})M_{\eta',n_2,n_3}({\bm k}_2,{\bm k}_3+{\bm G}+\delta{\bm G}),\nonumber\\
\end{eqnarray}
where $\delta'$ is the 2D periodic Kronecker delta function with period of primitive moir\'e reciprocal lattice vectors, $V({\bm q})$ is the Fourier transform of the interaction potential of holes, ${\bm G}$ is a reciprocal lattice vector, and $\delta{\bm G}={\bm k}_1+{\bm k}_2-{\bm k}_3-{\bm k}_4$. Here we adopt the gauge that $\psi_{\eta,n,{\bm k}_0+\bm G}({\bm r})=\psi_{\eta,n,{\bm k}_0}({\bm r})$ for any reciprocal lattice vector $\bm G$. Therefore, $u_{l,m_1,m_2}^{\eta,n,{\bm k}_0+\bm G'} = u_{l,m_1+m_1',m_2+m_2'}^{\eta,n,{\bm k}_0}$, where $\bm G' =m_1' \bm G_1+m_2' \bm G_2$.
In the following, we still choose $V({\bm q})=\frac{e^2}{4\pi\epsilon\epsilon_0}\frac{1}{\mathcal{A}}\frac{2\pi\tanh(|{\bm q}|d)}{|{\bm q}|}$, as in the FQH case. 
The matrix element of the impurity potential is 
\begin{eqnarray}
U^\eta_{\{{\bm k}_i\}\{n_i\}}=\sum_{\bm G}U({\bm k}_1-{\bm k}_2+{\bm G})M_{\eta,n_1,n_2}({\bm k}_1,{\bm k}_2-{\bm G}),\nonumber\\
\end{eqnarray}
where $U({\bm q})$ is the Fourier transform of the impurity potential. For numerical efficiency, one has to project Eq.~(\ref{Hh}) to a finite set of bands. The experimental observations of valley polarized FCIs at hole filling $\nu_h=2/3$ have been supported by single-band ED that only keeps the top valence band in a single valley~\cite{DiXiao_FCI_MoTe2,PhysRevB.108.085117}. ED calculations that keep both valleys and two bands per valley also indicate that the $\nu_h=2/3$ FCIs still survive in a small region of the $(\theta,\epsilon)$ parameter space~\cite{PhysRevB.109.045147}. 

\subsection{Density of holes in FCI ground states}
Before exploring quasiparticles, we first examine the real-space density of holes in the absence of quasiparticles for the valley polarized $\nu_h=2/3$ FCI ground states. The system size is labeled by the number of holes $N$ and the number of moir\'e unit cells $N_s$. For a general valley polarized ground state, the real-space density is 
\begin{eqnarray}
\rho({\bm r})=\sum_{\eta,n,n',{\bm k},{\bm k}'}\langle\gamma_{\eta, n,{\bm k}}^\dagger \gamma_{\eta, n',{\bm k}'}\rangle \psi^*_{\eta,n,{\bm k}}({\bm r}) \psi_{\eta,n',{\bm k}'}({\bm r}),\nonumber\\
\end{eqnarray}
where the expectation $\langle\cdot\rangle$ is over the ground state. The restriction ${\bm k}={\bm k}'$ is imposed if the ground state has translation symmetry. 
We project Eq.~(\ref{Hh}) to the top valence band of valley $K$, and calculate the density averaged over the three $\nu_h=2/3$ ground states at twist angle $\theta=3.7^\circ$. In this case, the real-space density can be simplified to $\rho({\bm r})=\sum_{\bm k} n_{\bm k} |\psi_{{\bm k}}({\bm r})|^2$, where $n_{\bm k}$ is the momentum-space occupation of holes and $\psi_{{\bm k}}({\bm r})$ is the wavefunction of the single band. To make finite-size systems as isotropic as possible, we apply ED to samples satisfying tilted geometry~\cite{Lauchli-PhysRevLett.111.126802,PhysRevB.90.245401}. The sample details used in this work are given in Appendix~\ref{app::tilted}.

\begin{figure}
\centerline{\includegraphics[width=\linewidth]{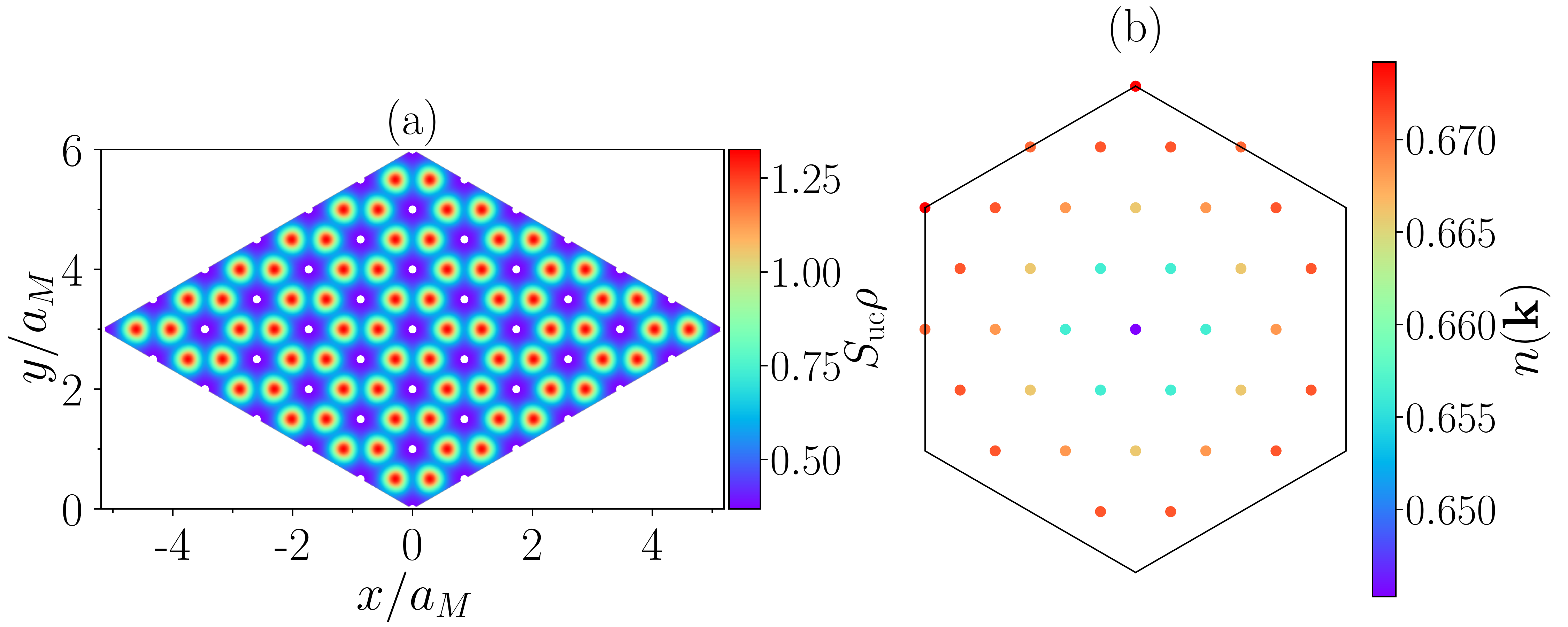}}
\caption{(a) The real-space density of holes for the $\nu_h=2/3$ FCI ground state in tMoTe$_2$. The system size is $N=24, N_s=36$. The white dots indicate the $\mathcal{R}_M^M$ positions. In (b), we show the momentum-space occupation (averaged over the three FCI ground states) for the system in (a). We set $\theta=3.7^\circ$, $d=10 \ {\rm nm}$ and $\epsilon=10$.}
\label{FCIgs}
\end{figure}

Unlike the FQH case, the real-space density of moir\'e FCIs is not uniform. In Fig.~\ref{FCIgs}(a), we show the data averaged over three nearly degenerate FCI ground states. $\rho({\bm r})$ is strongly peaked near the $\mathcal{R}_M^X$ and $\mathcal{R}_X^M$ positions, but almost vanishes near the $\mathcal{R}_M^M$ points. As the hole occupation in the momentum space is approximately uniform [Fig.~\ref{FCIgs}(b)], this non-uniform $\rho({\bm r})$ is determined by the real-space distribution of single-particle band wave functions. As shown in Ref.~\cite{PhysRevResearch.2.033087}, the holes in the top valence band indeed tend to be confined near the $\mathcal{R}_M^X$ and $\mathcal{R}_X^M$ positions when the first two valence bands carry Chern number $\mathcal{C}=\pm 1$, which holds for the parameters chosen by us. In this case, the first two valence bands in the $K$ valley can be approximated by the Haldane model on an effective  honeycomb lattice formed by $\mathcal{R}_M^X$ and $\mathcal{R}_X^M$ sites. We emphasize that the density non-uniformity only occurs in the length scale of one moir\'e unit cell for the FCI ground states. After summing the density within a moir\'e unit cell, we get the same total particle number $n_{\rm uc}=|\nu|/S_{\rm uc}$ for all unit cells, where $S_{\rm uc}$ is the area of a moir\'e unit cell. 

\begin{figure}
\centerline{\includegraphics[width=\linewidth]{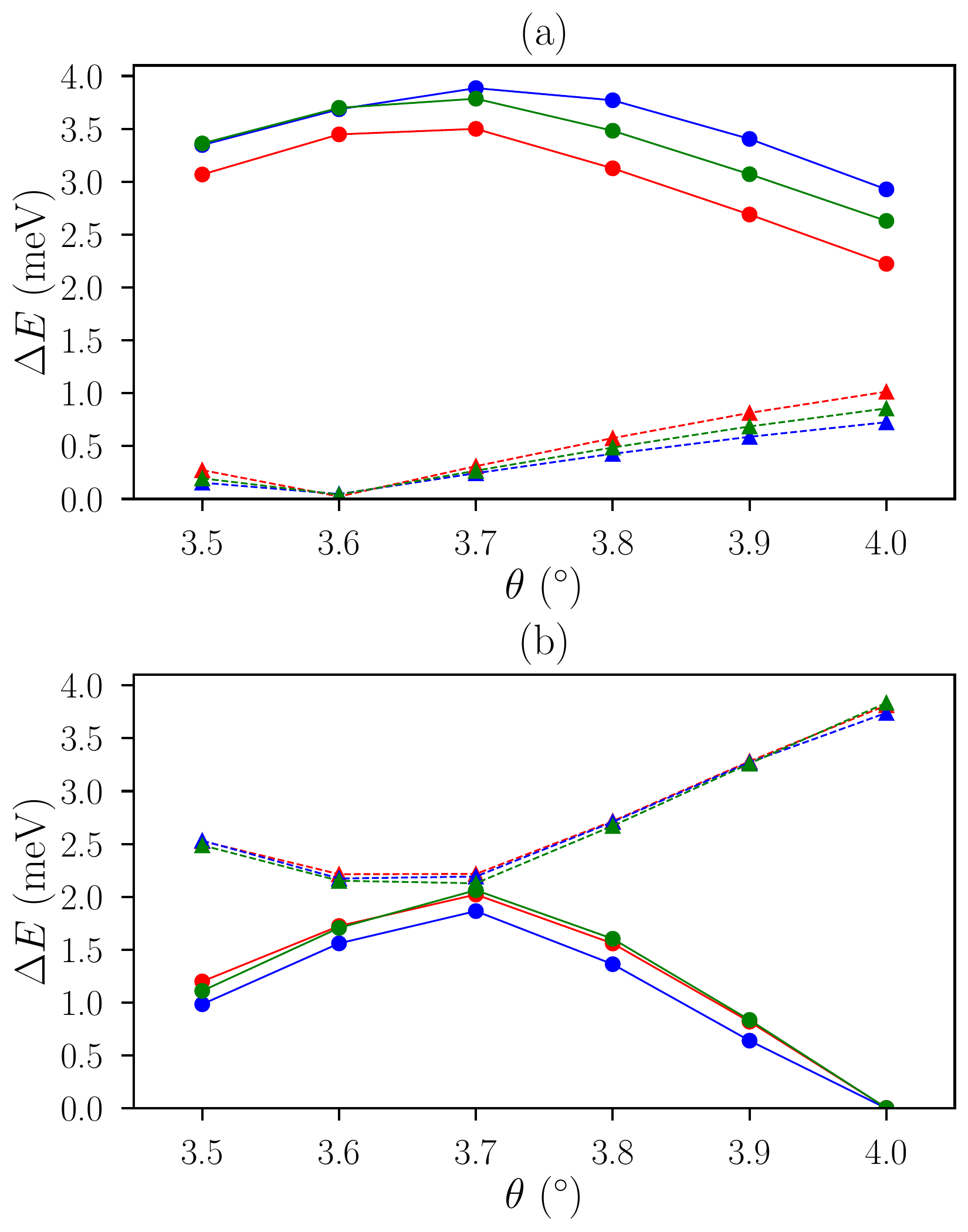}}
\caption{The energy gap (solid lines) and the splitting (dotted lines) of the quasiparticle manifold as a function of the twist angle. We consider the presence of a single delocalized quasiparticle and two delocalized quasiparticles in (a) and (b), respectively. The system sizes are $N=19, N_s=28$ (red), $N=21, N_s=31$ (blue), $N=23, N_s=34$ (green) in (a), and $N=20, N_s=29$ (red), $N=22, N_s=32$ (blue), $N=24, N_s=35$ (green) in (b).}
\label{qp_energy}
\end{figure}

\section{Quasiparticles in FCIs}
\label{sec:fci}
Now we study quasiparticles for the $\nu_h=2/3$ moir\'e FCIs in tMoTe$_2$. Note that the particle in ``quasiparticle'' actually means hole here. Like in the LLL, we require the number of moir\'e unit cells $N_s$ and the number of holes $N$ to satisfy $N_s=\frac{1}{2}(3N-N_{\rm qp})$ to generate $N_{\rm qp}$ quasiparticles.
It is natural to use a delta impurity potential $U_{\rm imp}({\bm w})$ of strength $U_0$ at position ${\bm w}$ that is attractive for holes to localize a quasiparticle at ${\bm w}$. The Fourier transform of the impurity potential is $U({\bm q})=\frac{U_0}{N_s}e^{-i{\bm q}\cdot{\bm w}}$. In this section, we choose $d=10 \ {\rm nm}$ and $\epsilon=10$ unless otherwise stated, and project the Hamiltonian to the top valence band of the $K$ valley. We will consider other parameter values and include more bands in Appendix~\ref{app::epsilon} and \ref{app::2B}. 

To reach larger system sizes, we still diagonalize the first two terms (band dispersion and interaction) in Eq.~(\ref{Hh}) to obtain states of delocalized quasiparticles, for which we have conserved momenta to reduce the many-body Hilbert space dimension. Then we diagonalize the impurity potentials in this subspace of mobile quasiparticles to pin the quasiparticles. Similar to the FQH case, the dimension of the quasiparticle subspace can be predicted by Haldane’s exclusion principle~\cite{Bernevig-2012PhysRevB.85.075128}. Naively, this strategy is expected to work well when $\Delta_s\ll U_0\ll \Delta_g$, such that the the impurity potential can sufficiently mix states within the quasiparticle subspace, but does not couple them to higher levels. Here $\Delta_g$ is the energy gap separating the subspace of delocalized quasiparticles from higher excited levels, and $\Delta_s$ is the energy splitting within the subspace. However, by diagonalizing the full Hamiltonian Eq.~(\ref{Hh}) with the single-band projection, we find that the mixing of the quasiparticle subspace with higher excited levels is still very weak even when the strength $U_0$ of the delta impurity potential has significantly exceeded $\Delta_g$ (see Appendix~\ref{app::fullED}).

In Fig.~\ref{qp_energy}, we display $\Delta_g$ and $\Delta_s$ in the presence of one and two delocalized quasiparticles at various twist angles, where $\Delta_g$ is defined as the energy difference between the highest level in the quasiparticle subspace and the lowest level out of this subspace. For a single delocalized quasiparticle, we have $\Delta_g>\Delta_s$ in the entire range of $\theta\in[3.5^\circ,4.0^\circ]$ [Fig.~\ref{qp_energy}(a)]. By contrast, the quasiparticle subspace is significantly broadened when two delocalized quasiparticles are present [Fig.~\ref{qp_energy}(b)]. Moreover, the gap is obviously smaller compared to the single quasiparticle case, reflecting the reduced charge gap in the presence of more quasiparticles. The broadening of the quasiparticle subspace and the shrinking of gap makes $\Delta_s>\Delta_g$ for two delocalized quasiparticles in the entire range of $\theta$ considered by us. In this case, the ratio $\Delta_s/\Delta_g$ is the smallest at $\theta=3.7^\circ$, which is the minimum (maximum) of $\Delta_s$ ($\Delta_g$) [Fig.~\ref{qp_energy}(b)]. We hence fix the twist angle at $\theta=3.7^\circ$ in the calculations throughout this section. 

\begin{figure}
\centerline{\includegraphics[width=\linewidth]{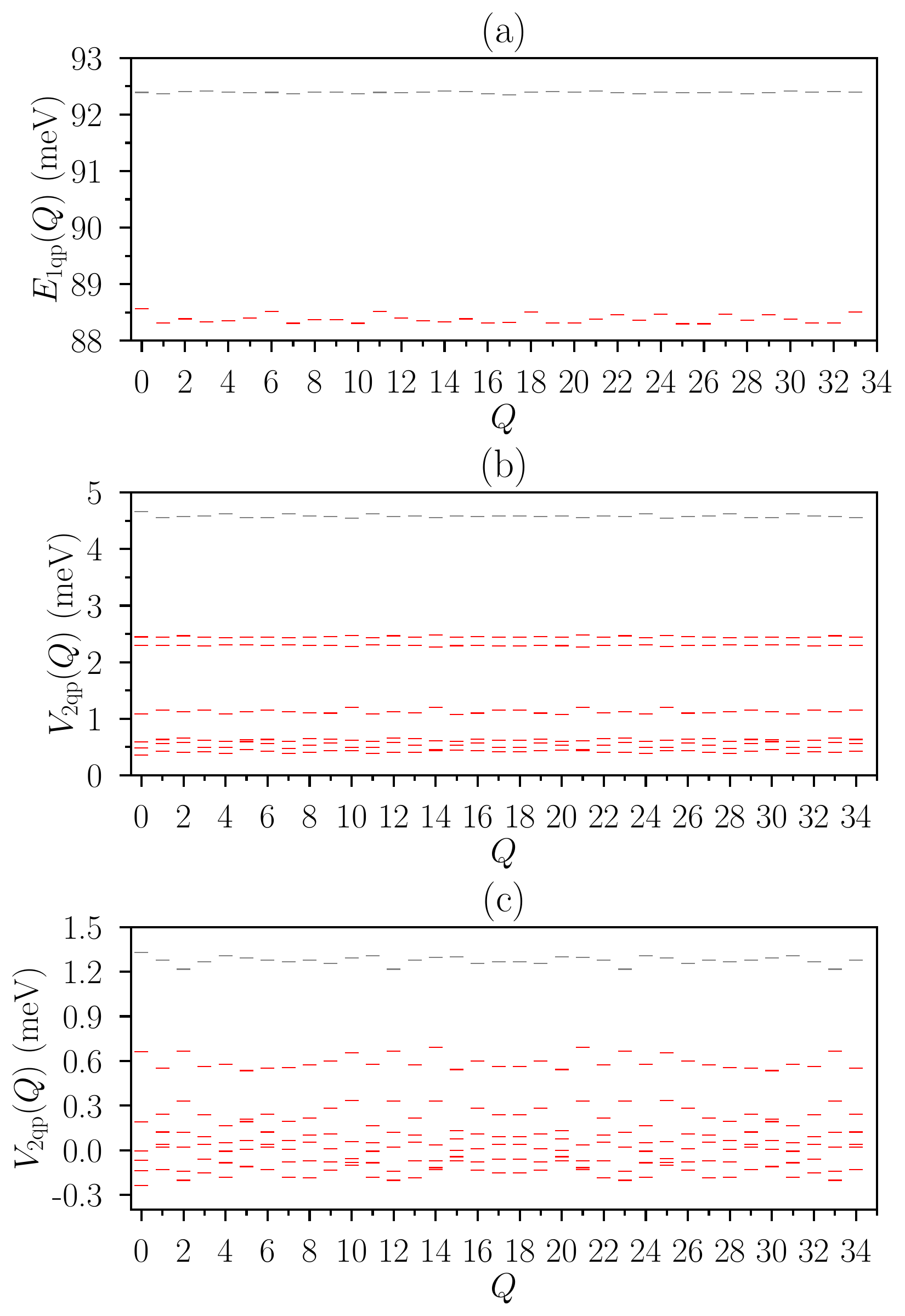}}
\caption{The low-energy energy spectra in presence of (a) a single delocalized quasiparticle and (b), (c) two delocalized quasiparticles. The system sizes are $N=23, N_s=34$ in (a) and $N=24, N_s=35$ in (b) and (c). $Q$ is a proper folding of the two-dimensional momentum ${\bm Q}$. The definitions of $E_{\rm 1qp}$ and $V_{\rm 2qp}$ are given in the text. Unlike in Ref.~\cite{goncalves2025spinlessspinfulchargeexcitations}, we do not shift the energies in (a) to make the dispersion near zero energy. The subspace of delocalized quasiparticles are highlighted by red. This subspace contains one level per momentum sector in (a), and six levels per momentum sector in (b) and (c). The gray levels are out of this subspace. We set $\theta=3.7^\circ$ and $\epsilon=10$. The sample-to-gate distance is chosen as $d=10 \ {\rm nm}$ in (b) and $d=2 \ {\rm nm}$ in (c).}
\label{qp_energy_K}
\end{figure}

\subsection{Single delocalized quasiparticles}
Using the energy levels in the presence of a single delocalized quasiparticle and the ground energy in the absence of quasiparticles, we can estimate the energy cost of a delocalized quasiparticle excitation. For a system of $N=N_0, N_s=N_{s,0}=3N_0/2$ without quasiparticles, there are three nearly degenerate FCI ground states with mean energy ${\bar E}_{\rm gs}$. We define the creation energy of a single delocalized quasiparticle as
\begin{eqnarray}
E_{\rm 1qp}({\bm Q})\equiv E_{N=N_0+1,N_s=N_{s,0}+1}({\bm Q})-{\bar E}_{\rm gs},
\end{eqnarray}
where $E_{N=N_0+1,N_s=N_{s,0}+1}({\bm Q})$ is the energy level with momentum ${\bm Q}$ of the many-body Hamiltonian $H$ (impurity potential excluded) of the system size $N=N_0+1,N_s=N_{s,0}+1$ (such that one delocalized quasiparticle is added to the ground state). We refer to Ref.~\cite{PhysRevB.90.245401} for the allowed momentum in tilted samples. In our calculations, we do not include the contribution from a uniform neutralizing charge background, which was considered by previous studies of FQH states in Landau levels~\cite{PhysRevLett.52.1583,PhysRevLett.54.237,PhysRevB.34.2670,PhysRevB.33.2221}. In Fig.~\ref{qp_energy_K}(a), we show $E_{\rm 1qp}({\bm Q})$ for $N_0=22$, whose ${\bar E}_{\rm gs}\approx 1.088 \ {\rm eV}$. $E_{\rm 1qp}({\bm Q})$ can also be understood as a dispersive band of the single quasiparticle, with $\Delta_s$ the bandwidth. This quasiparticle band is quite flat at all twisted angles considered by us. In usual Landau levels under a uniform magnetic field, the quasiparticle band is exactly flat. The weak dispersion observed here results from the non-uniform quantum geometry (Berry curvature and quantum metric) of the moir\'e band. Anyon dispersion in Landau levels under a non-uniform magnetic field (equivalent to non-uniform quantum geometry) was also reported recently~\cite{schleith2025anyondispersionnonuniformmagnetic}.

Analogous to Wannier states in a single-particle Bloch band, anyon Wannier states can be constructed for the single quasiparticle band. Since there is one state per momentum $\bm Q$ in the single quasiparticle band,  there is exactly one anyon Wannier state per unit cell in real space. The valley-polarized FCI state respect the $C_{3z}$ and $C_{2y}\mathcal{T}$ symmetries, where  $C_{nj}$ is the $n$-fold rotation around the $j$ axis and $\mathcal{T}$ is the time reversal symmetry. The symmetry-preserving anyon Wannier states are centered at the $\mathcal{R}_M^M$ sites with positions given by $\bm R= n_1 \bm a_1+ n_2 \bm a_2$, where $n_{1,2}$ are integers. 

\begin{figure}
\centerline{\includegraphics[width=\columnwidth]{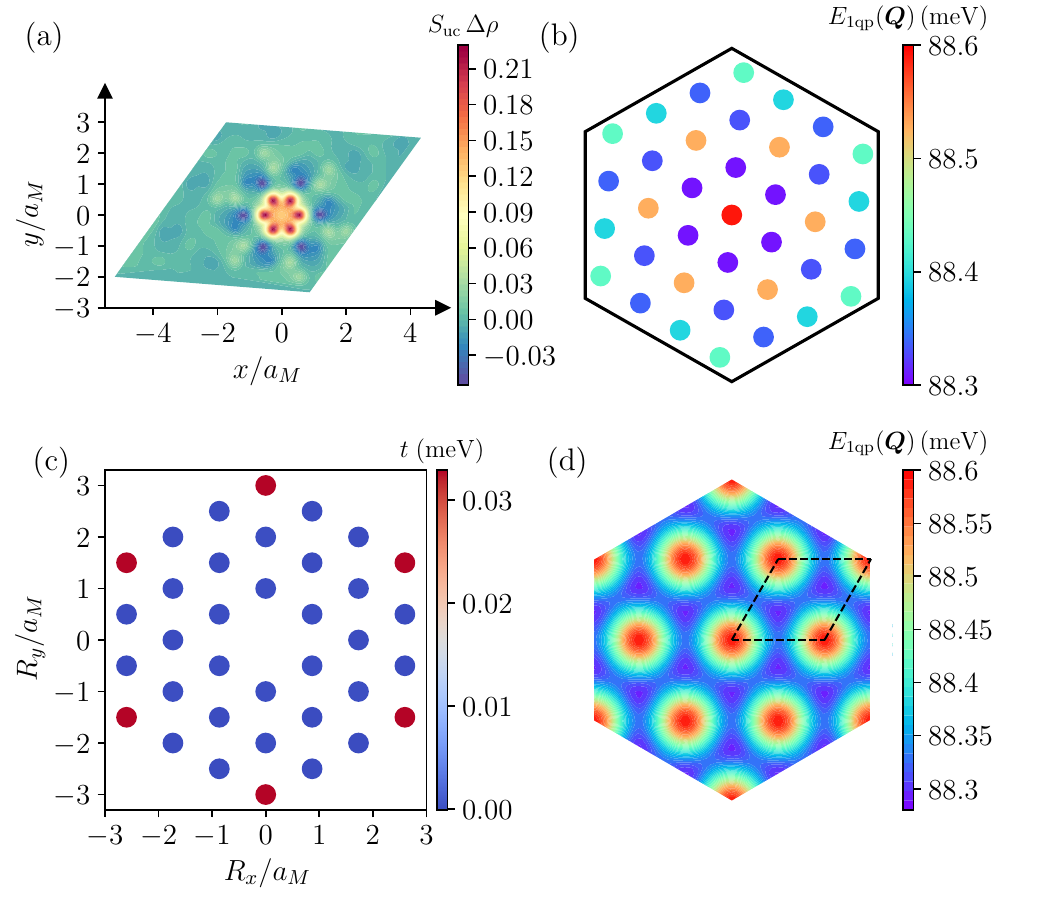}}
\caption{ 
(a) Density distribution of the anyon Wannier state centered at the origin measured relative to the FCI ground state. (b) The single quasiparticle energy $E_{\text{1qp}}(\bm Q)$ at discrete momenta $\bm Q$. (c) The anyon hopping parameter $t(\bm R)$ between the site at the origin and the site at position $\bm R$. (d) Interpolated $E_{\text{1qp}}(\bm Q)$ based on Eq.~\eqref{E1qinterpolation}. The black dashed lines mark the emergent periodicity in $E_{\text{1qp}}(\bm Q)$. The system size is $N=25$ and $N_s=37$. 
}
\label{anyonTB}
\end{figure}

An anyon Wannier state $|\bm R \rangle_{\text{1qp}}$ centered at $\bm R$ is a linear superposition of the single quasiparticle states $|\Phi_{\bm Q}\rangle_{\text{1qp}}$ at different momenta $\bm Q$,
\begin{equation}
    |\bm R \rangle_{\text{1qp}}= \frac{1}{\sqrt{N_s}} \sum_{j=0}^{N_s-1}  e^{i \phi_j} e^{-i\bm Q_j \cdot \bm R} |\Phi_{\bm Q_j}\rangle_{\text{1qp}}.
\end{equation}
For numerically returned states $|\Phi_{\bm Q}\rangle_{\text{1qp}}$ in the subspace of one mobile quasiparticle, we determine the phase factors $ e^{i \phi_j}$ as follows.
Without loss of generality, we take $\phi_0=0$. The remaining phases $\phi_{j\neq 0}$ are fixed by requiring that the matrix element $ {}_{\text{1qp}} \langle \Phi_{\bm 0} | \Delta(\bm r) e^{i \phi_j}|\Phi_{\bm Q_j}\rangle_{\text{1qp}}$ is real and positive, where $\Delta(\bm r)= + \sum_{i=1}^{N} \delta(\bm r_i)$.  This gauge choice maximizes the density at the origin for the Wannier state $|\bm R= \bm 0 \rangle_{\text{1qp}}$  with respect to $\phi_j$. A generic Wannier state $|\bm R' \neq \bm 0 \rangle_{\text{1qp}}$ is related to $|\bm R= \bm 0 \rangle_{\text{1qp}}$  by translating all particles by a lattice vector $\bm R'$. To illustrate the resulting states, we plot the density difference $\Delta \rho$ between the anyon Wannier state centered at $\bm R= \bm 0$  and the FCI ground state, as illustrated in Fig.~\ref{anyonTB}(a). As expected, the profile of $\Delta \rho$ is centered at the origin and decays to zero for distance far away from the center. However, $\Delta \rho$ peaks at the six $\mathcal{R}_M^X$ and $\mathcal{R}_X^M$ sites around the origin, following the density distribution in the ground state. 

We further calculate the effective anyon hopping parameter $t(\bm R)$ between two $\mathcal{R}_M^M$ sites separated by $\bm R$,
\begin{equation}
    t(\bm R) = {}_{\text{1qp}} \langle \bm R | H | \bm 0 \rangle_{\text{1qp}} =  \frac{1}{N_s} \sum_{\bm Q} e^{i \bm Q \cdot \bm R} E_{\text{1qp}} (\bm Q).
\label{tR}
\end{equation}
$E_{\text{1qp}} (\bm Q)$, the quasiparticle energy measured relative to ${\bar E}_{\rm gs}$, is plotted in Fig.~\ref{anyonTB}(b) at discrete momenta for the system size $N=25,N_s=37$.  Based on these values of $E_{\text{1qp}} (\bm Q)$, the calculated $t(\bm R)$ is illustrated in Fig.~\ref{anyonTB}(c). Due to the discrete nature of the Fourier transform in Eq.~\eqref{tR}, we obtain $36$ independent hopping parameters $t(\bm R \neq \bm 0)$ in addition to the onsite energy $t(\bm 0)$ for this system size. An intriguing observation is that the dominant hopping amplitudes occur between sites separated by a distance of three moir\'e lattice constants. For comparison, $t(|\bm R|=3a_M) \approx 0.033 $ meV, which is at least two orders of magnitude larger than all other hopping parameters in Fig.~\ref{anyonTB}(c). 

Having obtained $t(\bm R)$ using $E_{\text{1qp}} (\bm Q)$ of a finite-size system, we can evaluate $E_{\text{1qp}} (\bm Q)$ in the thermodynamic limit via the inverse Fourier transform
\begin{equation}
    E_{\text{1qp}} (\bm Q) =\sum_{\bm R} e^{-i \bm Q \cdot \bm R} t(\bm R),
\label{E1qinterpolation}
\end{equation}
where $\bm Q$ is no longer restricted to the discrete momenta imposed by the finite system size. Fig.~\ref{anyonTB}(d) shows $E_{\text{1qp}} (\bm Q)$ as a function of continuous $\bm Q$, which exactly reproduces the discrete values shown in Fig.~\ref{anyonTB}(b). Interestingly, $E_{\text{1qp}} (\bm Q)$ is periodic not only  with respect to the moir\'e Brillouin zone defined by the underlying lattice, but also with respect to a smaller effective Brillouin zone, delineated by the dashed lines in Fig.~\ref{anyonTB}(d). This emergent periodicity arises because $t(\bm R)$ is dominated by hopping between sites separated by $3 a_M$, leading to an effective Brillouin zone that is $3\times 3$ smaller than the original one. We note that this emergent periodicity in $E_{\text{1qp}} (\bm Q)$ has been theoretically anticipated in Ref.~\cite{Shi2025Doping} and  independently reported in numerical results of  Ref.~\cite{goncalves2025spinlessspinfulchargeexcitations}. Here we provide a perspective based on the anyon Wannier states and the hopping model. 

We compare the interpolated $E_{\text{1qp}} (\bm Q)$ in Fig.~\ref{anyonTB}(d)  to the values in Fig.~\ref{qp_energy_K}(a) at the discrete momenta $\bm Q$ allowed in a different system size of $N=23,N_s=34$. The energy differences at each momentum can be as small as $3\times 10^{-4}$ meV and do not exceed $0.026$ meV. This close agreement provides a benchmark for the accuracy of the anyon hopping model in capturing the quasiparticle energy dispersion, with hopping parameters truncated to $|\bm R| \leq 3a_M$.

\subsection{Two delocalized quasiparticles}
Given the nearly flat single-quasiparticle band, the significant broadening of the subspace of two delocalized quasiparticles is a signal of the interaction between quasiparticles. For finite systems it is difficult to precisely subtract the contribution of single-quasiparticle dispersion from the two-quasiparticle spectrum, as their allowed momentum points are different. Here we only make a rough estimation of the interaction between two quasiparticles by calculating 
\begin{eqnarray}
V_{\rm 2qp}({\bm Q})\equiv E_{N=N_0+2,N_s=N_{s,0}+2}({\bm Q})-{\bar E}_{\rm gs}-2{\bar E}_{\rm 1qp},\nonumber\\
\end{eqnarray}
where $E_{N=N_0+2,N_s=N_{s,0}+2}({\bm Q})$ is the energy level with momentum ${\bm Q}$ of the many-body Hamiltonian of the system size $N=N_0+2,N_s=N_{s,0}+2$ (such that two delocalized quasiparticles are added to the ground state), and ${\bar E}_{\rm 1qp}$ is the $E_{\rm 1qp}({\bm Q})$ averaged over all momentum sectors. The result for $N_0=22$ is displayed in Fig.~\ref{qp_energy_K}(b), whose ${\bar E}_{\rm 1qp}\approx 0.0884 \ {\rm eV}$. Remarkably, $V_{\rm 2qp}$ is positive in all momentum sectors, with the lowest value $\approx 0.35 \ {\rm meV}$. As the neglected single-quasiparticle bandwidth $\Delta_s$ is only $\approx0.27 \ {\rm meV}$ in this case, it is very likely that the two delocalized quasiparticles have a repulsive interaction. 

The interactions between quasiparticles and quasiholes have been studied in the context of Landau levels~\cite{PhysRevB.53.9599,PhysRevB.61.2846,PhysRevB.109.L201123,trung2025longrangeentanglementrolerealistic,xu2025dynamicsclustersanyonsfractional}. We do similar analysis for the $\nu=2/3$ FQH quasiparticles in our LLL setup in Sec.~\ref{sec:continuum}, using the same system sizes as in Fig.~\ref{qp_energy_K}. In that case, we also find repulsive interactions between two quasiparticles. Motivated by a recent work which reported the crossover from repulsive to attractive interaction between two $\nu=1/3$ Laughlin quasiholes when the range of the electron-electron interaction is decreased~\cite{xu2025dynamicsclustersanyonsfractional}, we have also examined much smaller $d$ ($d=2 \ {\rm nm}$ for tMoTe$_2$ and $d=\ell_B$ for the LLL), corresponding to stronger screening. Indeed, we find negative levels in $V_{\rm 2qp}({\bm Q})$, which is a signature of attractive interaction between quasiparticles [see Fig.~\ref{qp_energy_K}(c) for the data in tMoTe$_2$]. Therefore, our results in both the LLL and tMoTe$_2$ are consistent with the observation in Ref.~\cite{xu2025dynamicsclustersanyonsfractional}.  

In the following, we include impurity potentials and characterize the localized $\nu_h=2/3$ FCI quasiparticles in tMoTe$_2$.

\begin{figure*}
\centerline{\includegraphics[width=\linewidth]{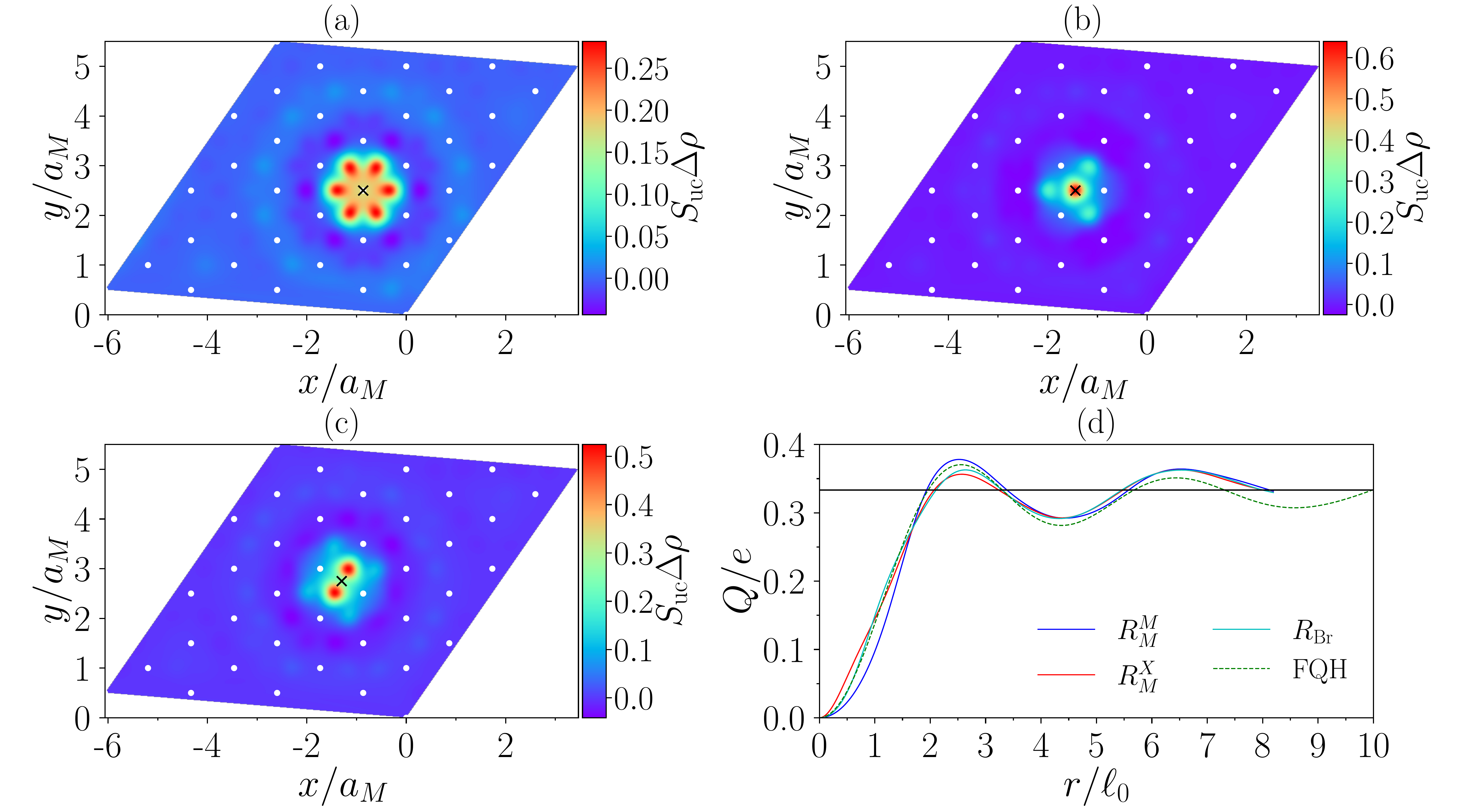}}
\caption{(a)-(c) Density distribution with a localized single-quasiparticle excitation in the $\nu_h=2/3$ tMoTe$_2$ FCI with $N=25,N_s=37$. The impurity potential (indicated by the black cross) is located at (a) $\mathcal{R}_M^M$, (b) $\mathcal{R}_M^X$, and (c) $\mathcal{R}_{\rm Br}$, respectively. (d) Excess charges corresponding to the three cases in (a)-(c). We also give the data of the $\nu=2/3$ Coulomb FQH state [dashed line, the same as in Fig.~\ref{qh1qp}(b)] for comparison. The horizontal reference line  indicates $Q=e/3$, which is expected for a Laughlin-type quasiparticle.}
\label{FCI_1qp}
\end{figure*}

\subsection{Single localized quasiparticle}
\label{sec:fci_1qp}
Let us first consider a single localized quasiparticle of the $\nu_h=2/3$ FCIs in tMoTe$_2$. Again we get three nearly degenerate ground states after diagonalizing the impurity potential in the subspace of a mobile quasiparticle. We are interested in the effect of this localized quasiparticle on particle's real-space density, which is measured by 
\begin{eqnarray}
\Delta\rho({\bm r})\equiv\rho({\bm r})-\rho_0({\bm r}).
\end{eqnarray}
Here $\rho({\bm r})$ is the particle's density in the presence of a single quasiparticle (averaged over the three localized quasiparticle states), and $\rho_0({\bm r})$ is the averaged particle's density in the three FCI ground states without quasiparticles [Fig.~\ref{FCIgs}(a)]. In Figs.~\ref{FCI_1qp}(a), \ref{FCI_1qp}(b), and \ref{FCI_1qp}(c), we show $\Delta\rho({\bm r})$ for $N=25,N_s=37$, which is the largest system size we can reach by ED. The impurity potential is located at $\mathcal{R}_M^M$, $\mathcal{R}_M^X$, and the bridge point $\mathcal{R}_{\rm Br}$ (middle of $\mathcal{R}_M^X$ and $\mathcal{R}_X^M$), respectively. Extra particles indeed concentrate around the impurity potential in all of the three cases, which is a signal of a localized quasiparticle. With increasing distance from the impurity potential, $\Delta\rho({\bm r})$ oscillates and tends to zero. Unlike the FQH case [Fig.~\ref{qh1qp}(a)], the density distribution in the presence of an FCI quasiparticle no longer has continuous rotational symmetry. Moreover, there is a clear dependence of $\Delta\rho({\bm r})$ on the position of the impurity potential. One can see that particles still prefer to occupy $\mathcal{R}_M^X$ and $\mathcal{R}_X^M$ regions as in the ground-state case, such that $\Delta\rho({\bm r})$ demonstrates $C_{6z}$ and $C_{3z}$ rotation symmetry, respectively, if the impurity potential is located at $\mathcal{R}_M^M$ and $\mathcal{R}_M^X$ sites [Figs.~\ref{FCI_1qp}(a) and \ref{FCI_1qp}(b)]. The density profile for the impurity located at $\mathcal{R}_M^M$ closely follows that of the anyon Wannier state shown in Fig.~\ref{anyonTB}(a).

To further characterize the existence of a localized quasiparticle, we compute the charge excess $Q(r)$ as a function of the distance $r$ from the impurity potential, as defined in Eq.~(\ref{eq::Qr}). Note that in the FCI case both $\rho({\bm r})$ and $\rho_0({\bm r})$ are not uniform or rotationally invariant, and we cannot reduce Eq.~(\ref{eq::Qr}) to an integral in the radial direction. The results corresponding to Figs.~\ref{FCI_1qp}(a), \ref{FCI_1qp}(b), and \ref{FCI_1qp}(c) are similar and we display them in Fig.~\ref{FCI_1qp}(d). For direct comparison with the FQH data, we choose the effective magnetic length~\cite{PhysRevB.91.045126}
\begin{eqnarray}
\ell_0=\left(\frac{\sqrt{3}}{4\pi}\right)^{1/2}a_M
\end{eqnarray}
as the length unit for tMoTe$_2$. Under this choice, $d=10 \ {\rm nm}$ corresponds to $d/\ell_0\approx 5$, which is consistent with the setting $d=5\ell_B$ in Sec.~\ref{sec:continuum}. When $r\gtrsim 2\ell_0$, one can see oscillations of the excess charge around $e/3$ -- the theoretical value of a Laughlin-type quasiparticle. This is a clear signal that the quasiparticle starts nucleating. However, $Q(r)$ does not converge even at the largest $r$ we can reach, suggesting that our system size is still insufficient for the quasiparticle to fully develop. The excess charge of $\nu_h=2/3$ moir\'e FCIs is similar to that of the $\nu=2/3$ FQH state when the length units are chosen as $\ell_0$ and $\ell_B$ for tMoTe$_2$ and the LLL, respectively, but the oscillations of $Q(r)$ in the FQH case show quicker damping. If we take the radius of the $\nu=2/3$ FQH quasiparticle of Coulomb interaction as $6\ell_B$ as suggested by the braiding data in Sec.~\ref{continuumb}, the $\nu_h=2/3$ FCI quasiparticle in tMoTe$_2$ extents within a circle whose radius is at least $6\ell_0\approx 2.2a_M\approx 120 \ \mathring{\mathrm A}$. This spatial extent could further increase if multiple bands are taken into account, as shown in Appendix~\ref{app::2B}.

\begin{figure}
\centerline{\includegraphics[width=\linewidth]{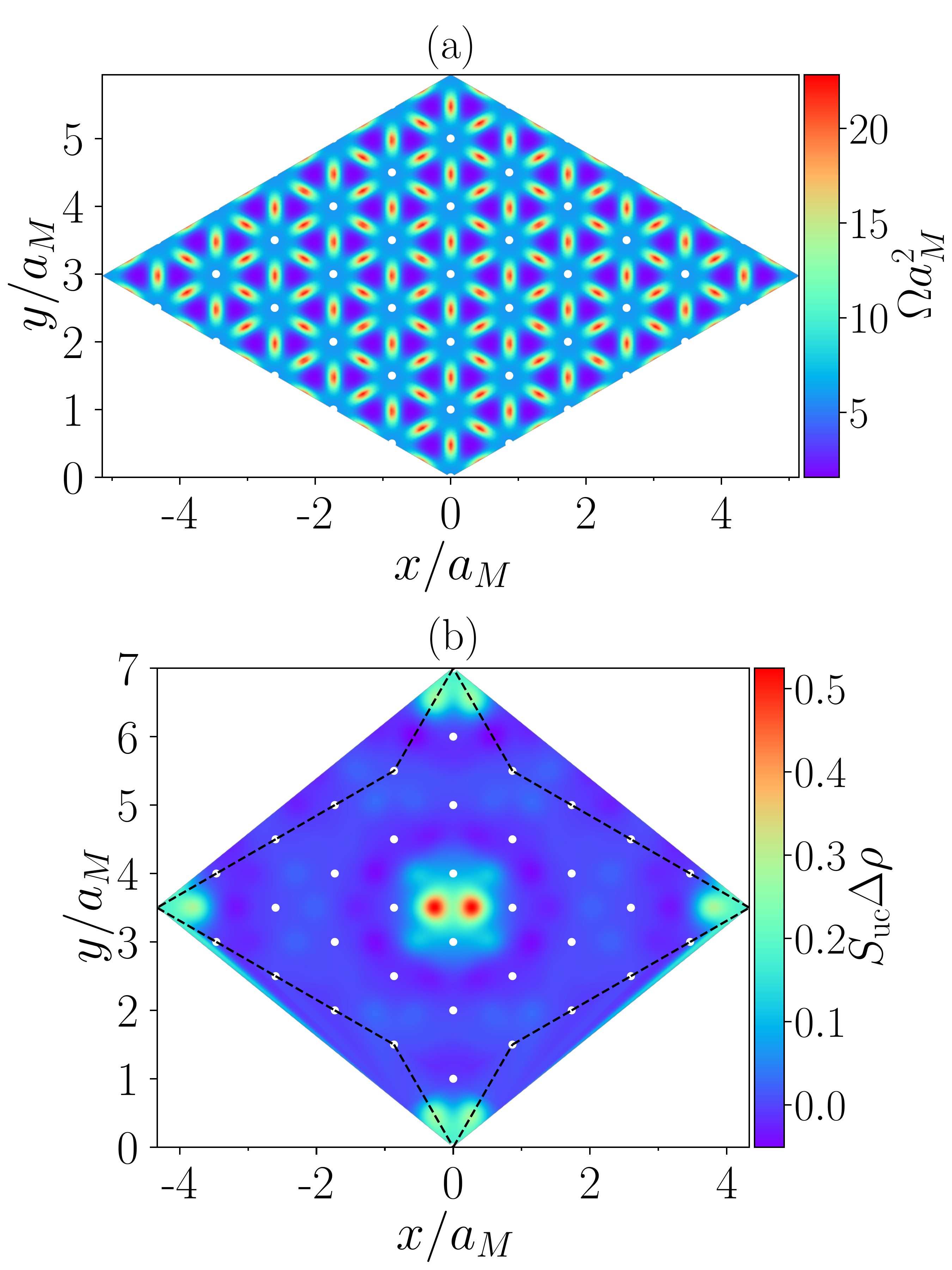}}
\caption{(a) The real-space Berry curvature probed by dragging a hole residing in the $K$-valley top valence band. (b) Braiding path (dashed line) of two quasiparticles for the system size $N=24, N_s=35$. The density distribution shown in the plot corresponds to two impurity potentials at the sample center and ${\bm r}=(0,0)$. We have averaged the density over the three nearly degenerate states with two localized quasiparticles.}
\label{FCI_2qp}
\end{figure}

\subsection{Braiding of two quasiparticles}
We now examine the fractional statistics of quasiparticle excitations in the $\nu_h=2/3$ tMoTe$_2$ moir\'e FCIs. We should emphasize that, while there is no physical magnetic field in moir\'e FCIs, a real-space Berry curvature $\Omega({\bm r})$ does exist, which contributes to the Berry phase when two localized quasiparticles are braided. This real-space Berry curvature has been reported for tight-binding models~\cite{PhysRevB.91.045126}. For tMoTe$_2$, one can demonstrate $\Omega({\bm r})$ by mapping the tMoTe$_2$ model to a Landau level problem, in which an effective periodic magnetic field with nonzero mean emerges~\cite{PhysRevLett.132.096602}. To numerically probe this real-space Berry curvature, we adiabatically drag a single hole using a mobile impurity potential projected to the top valence band in the $K$ valley. If the hole's wave function accumulates a Berry phase $\Phi({\bm r})$ along a closed path around point ${\bm r}$, we can approximate the real-space Berry curvature at ${\bm r}$ as 
\begin{eqnarray}
\Omega({\bm r})=\lim_{S({\bm r})\rightarrow 0}\Phi({\bm r})/S({\bm r}),
\end{eqnarray}
where $S({\bm r})$ is the area enclosed by the closed path. In Fig.~\ref{FCI_2qp}(a), we show $\Omega({\bm r})$ for the $K$-valley top valence band of tMoTe$_2$ at twist angle $\theta=3.7^\circ$. It is indeed very similar to the effective magnetic field obtained in Ref.~\cite{PhysRevLett.132.096602}. For a single hole, the accumulated Berry phase along the boundary of a moir\'e unit cell is $2\pi$. Moreover, dragging a single localized quasiparticle along the same path gives Berry phase $2\pi/3$, which provides another compelling evidence of the fractional charge $e/3$ of the quasiparticle. 

Then we fix one impurity potential at the sample center and move the other along a closed path, as shown in Fig.~\ref{FCI_2qp}(b). We carefully choose the path such that the two impurity potentials are separated as far as possible and the path encloses an integer number of moir\'e unit cells. The technical method used to extract the Berry phase is the same as that in Sec.~\ref{continuumb}. The accumulated Berry phase still contains two parts: the AB phase resulting from the real-space Berry curvature $\Omega({\bm r})$, and the anyon statistical phase. The AB phase is just $2\pi/3$ times the number of enclosed moir\'e unit cells. For the path in Fig.~\ref{FCI_2qp}(b), we numerically find the anyon statistical phase as $1.311\pi$ for $N=24, N_s=35$, which has reasonably good agreement with the theoretical prediction $4\pi/3$. Again, the discrepancy is because the quasiparticles are not fully developed and separated in the finite system.

\section{Trial Wave function for anyons in FCIs}
\label{sec::motecarlo}
We now reveal the connection between the FQH states in the LLL and the FCIs in tMoTe$_2$ regarding the quasiparticles. The $\nu=1/3$ Laughlin wave function in the LLL is
\begin{align}
\Psi_L = \mathcal{N}_L \prod_{1\le i<j\le N} (z_i - z_j)^3 e^{-\frac{1}{4\ell_B^2}\sum_i |z_i|^2}, 
\label{eq:Laughlin}
\end{align}
where $z_j = x_{j} + i y_{j}$ is the complex coordinate of particle $j$, $N$ is the number of particles, and $\mathcal{N}_L$ is a normalization factor. Here we write the wave functions on the disk geometry for symbol simplicity. The quasihole excitation on top of the Laughlin state is given by
\begin{align}
\Psi_{L,h} = \mathcal{N}_{L,h} \prod_i (z_i-w) \Psi_L, 
\label{eq:Laughlinqh}
\end{align}
where $w$ is the complex coordinate of the quasihole center and $\mathcal{N}_{L,h}$ is a normalization factor. Both $\Psi_L$ and $\Psi_{L,h}$ are zero-energy eigenstates of the first-order Haldane's pseudopotential interaction~\cite{PhysRevLett.51.605}.   

It has been observed that the top moir\'e valence band of tMoTe$_2$ is similar to the LLL, as it nearly saturates the trace inequality $T \ge 0$ of the single-particle quantum geometry~\cite{PhysRevResearch.5.L032022,PhysRevLett.132.096602,ChiralMoTe2}. Here $T$ is defined as~\cite{PhysRevB.90.165139,GeometryJackson,liu2024theorygeneralizedlandaulevels}
\begin{equation}
    T=\min_{\tilde g}\left(\frac{1}{2\pi}\int_{\text{MBZ}} {\rm tr}_{\tilde g} g({\bm k})\right) - |\mathcal{C}|
\end{equation}
where $g({\bm k})$ is the quantum metric of the band, and ${\rm tr}_{\tilde g} g\equiv\sum_{a,b=x,y}{\tilde g}_{ab}g^{ab}$ is the generalized trace with respect to a unimodular matrix $\tilde g$.  For the top moir\'e valence band in tMoTe$_2$, $T$ is approximately $0.1 \sim 0.3$, as discussed in detail in Appendix~\ref{app::2B}. In comparison, the LLL has exactly $T=0$ and momentum-independent quantum geometric tensor. More generally, an ideal Chern band with $T=0$ and momentum-dependent quantum geometries can be viewed as a generalized LLL with the wave functions of the form 
\begin{eqnarray}
\Theta_{\boldsymbol k}(\boldsymbol r)=\mathcal{N}_{\boldsymbol{k}}\mathcal B(\boldsymbol r)\Psi_{0,\boldsymbol k} (\boldsymbol{r}),
\end{eqnarray}
where $\boldsymbol{k}$ is the wave vector in the Brillouin zone, $\Psi_{0,\boldsymbol k} (\boldsymbol{r})$ is the magnetic Bloch wave function of the LLL, $\mathcal{B}(\boldsymbol r)$ is a $\boldsymbol{k}$-independent but spatially-varying function, and  $\mathcal{N}_{\boldsymbol{k}}$ is a normalization factor~\cite{Ledwith2020Fractional,Wang2021Exact,PhysRevB.108.205144}. Since $T$ is small in tMoTe$_2$, the Bloch wave function $\psi_{\bm{k}}(\bm{r})$ for its top moir\'e valence band in $K$ valley can be approximated by $\Theta_{\boldsymbol k}(\boldsymbol r)$. We obtain the function $\mathcal B(\boldsymbol r)$ by maximizing the momentum-averaged value of the overlap $|\langle \psi_{\bm{k}} | \Theta_{\boldsymbol k} \rangle|^2$ using a variational approach as outlined in Ref.~\cite{Li2025Variational}. The function $\mathcal B(\boldsymbol r)$ captures the density fluctuations in the moir\'e band, and $|\mathcal B(\boldsymbol r)|^2$ peaks at $\mathcal{R}_M^X$ and $\mathcal{R}_X^M$ positions similar to the density shown in Fig.~\ref{FCIgs}(a).

\begin{figure}
\centerline{\includegraphics[width=1\linewidth]{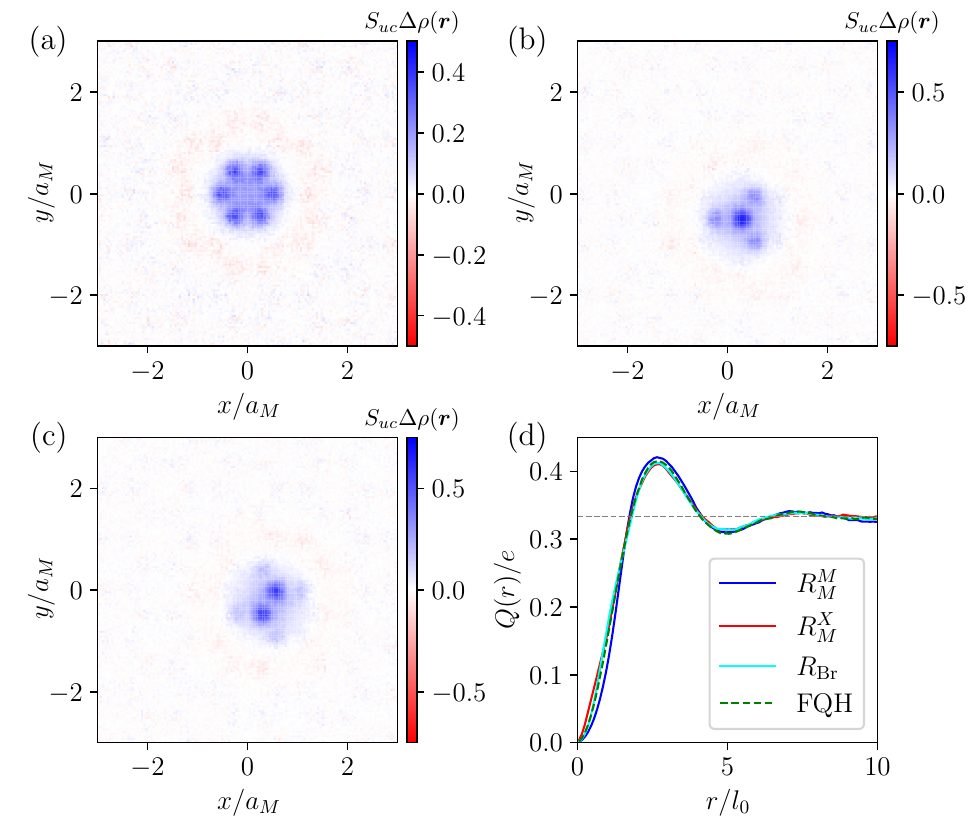}}
\caption{(a)-(c) Monte Carlo results of $\Delta \rho(\bm r)$ using Eq.~\eqref{PhiLLh}, with a single quasiparticle located at (a) $\mathcal{R}_M^M$, (b) $\mathcal{R}_M^X$, and (c) $\mathcal{R}_{\rm Br}$, respectively. (d) Excess charges corresponding to the three cases in (a)-(c). As a comparison, we plot the excess charge for the LLL using green dashed line (labeled by FQH), which is also obtained using the Monte Carlo simulation. The expected quasiparticle charge $Q = e/3$ is plotted in horizontal dashed line. }
\label{MC}
\end{figure}

Both the $\nu=1/3$ Laughlin model wave function in the LLL and its quasihole excitation can be generalized to FCIs in a Chern band with the single-particle wave function $\Theta_{\boldsymbol k}(\boldsymbol r)$:
\begin{equation}
\begin{aligned}
\Phi_{L} &= \mathcal{N}_{L}^{'} \prod_i  \mathcal B(\boldsymbol r_i) \Psi_L, \\
\Phi_{L,h} & = \mathcal{N}_{L,h}^{'} \prod_i (z_i-w) \mathcal B(\boldsymbol r_i) \Psi_L, 
\end{aligned}  
\label{PhiLLh}
\end{equation}
where $\mathcal{N}_{L}^{'} $ and $\mathcal{N}_{L,h}^{'}$ are normalization factors. Here $\Phi_{L}$ and $\Phi_{L,h}$ are trial wave functions for, respectively, the FCI ground state and its quasihole excitation at filling $\nu=1/3$. By performing particle-hole transformation to $\Phi_{L}$ ($\Phi_{L,h}$) within the manifold spanned by $\Theta_{\boldsymbol k}(\boldsymbol r)$, we obtain, respectively, the FCI ground state (quasiparticle excitation) at filling $\nu=2/3$. Therefore, the density distribution $\Delta\rho(\bm r)$ of the FCI quasiparticle at $\nu=2/3$ can be determined by the density difference between  $\Phi_{L}$ and $\Phi_{L,h}$. 

With the optimal $\mathcal B(\boldsymbol r)$ for the $K$-valley top valence band of tMoTe$_2$, we separately calculate the density of $\Phi_{L}$ and $\Phi_{L,h}$ using classical Monte Carlo simulation by sampling $3 \times 10^6$ configurations (after $10^6$ burn-in steps) for a system of $N=64$ particles, which is far beyond the capability of ED. 
Figs.~\ref{MC}(a)-\ref{MC}(c) show the Monte Carlo results of $\Delta\rho(\bm r)$ with $w$ located at $\mathcal{R}_M^M$, $\mathcal{R}_M^X$, and $\mathcal{R}_{\text{Br}}$, respectively. The quasiparticle density distribution depends on $w$, capturing all the qualitative features in the ED results shown in Fig.~\ref{FCI_1qp}. The corresponding excess charges $Q(r)$ of the quasiparticle as functions of distance $r$ from $w$ are shown in Fig.~\ref{MC}(d). Notably, these $Q(r)$ curves, with minor differences at short distances, are all similar with each other, which is consistent with that shown in Fig.~\ref{FCI_1qp}(d). Based on these comparisons, we conclude that the particle-hole conjugate of $\Phi_{L,h}$ in Eq.~\eqref{PhiLLh} provides a reasonable trial wave function for a localized $\nu_h=2/3$ FCI quasiparticle in a single top valence band of tMoTe$_2$ and captures the real-space density distribution. Note that $Q(r)$ obtained from Eq.~\eqref{PhiLLh} converges faster to $e/3$ than those in Fig.~\ref{FCI_1qp}(d), thus $\Phi_{L,h}$ underestimates the spatial extend of the quasiparticle of Coulomb states. In Fig.~\ref{MC}(d), we also plot the Monte Carlo results for $Q(r)$ in the LLL using the wave functions $\Psi_L$ and $\Psi_{L,h}$, which  
matches quantitatively well with the first-order Haldane's pseudopotential quasiparticle result in Fig.~\ref{qh1qp}(b). Moreover, it is very close to the other three $Q(r)$ curves with nontrivial $\mathcal B(\boldsymbol r)$. This observation again indicates that the dominant effect of $\mathcal B(\boldsymbol r)$ is the modulation of charge density only within the scale of a lattice unit cell. 

\section{Conclusion}
\label{sec:conclusion}
In this work, we have investigated the properties of quasiparticles for the $\nu_h=2/3$ FCI in tMoTe$_2$ and for the $\nu=2/3$ FQH state in the LLL. We have proposed a tight-binding model for mobile FCI quasiparticles. We further pin quasiparticles by delta impurity potentials. We get clear signal of the nucleation of localized quasiparticles, as shown by the consistency of numerically obtained charge and braiding phase of quasiparticles with theoretical values. The spatial extent of a localized FQH quasiparticle is estimated by either using the density profile or the braiding phase. The size of a FCI quasiparticle can then be estimated by mapping to its FQH counterpart via an effective lattice magnetic length. Trial wavefunctions of FCI quasiparticles are variationally determined using the nearly ideal quantum geometry of the tMoTe$_2$ top valence band. The effective interaction between two $e/3$ FCI quasiparticles shows a crossover from repulsion to attraction when the gate screening in the Coulomb potential is enhanced, which may provide a way to form $2e/3$ anyons.

There are several possible future directions based on this work. First, our numerical characterization of FQH/FCI quasiparticles still suffers from non-negligible finite-size effects. Even within the single-band/LLL projection, the localized quasiparticle does not fully develop, as reflected by the oscillating excess charge and residual deviation of the braiding phase from the theoretical value. The $\nu=2/3$ quasihole is even much bigger, thus we cannot extract useful information about it by ED (see Appendix~\ref{app::qh}). We have also made tentative investigations of band mixing effects on FCI quasiparticles in tMoTe$_2$ (see Appendix~\ref{app::2B}), but the results are limited to smaller systems and much less conclusive. More advanced numerical techniques are hence necessary for solving these problems. If the recent applications of neural-network algorithms to FQH and FCI systems~\cite{PhysRevLett.134.176503,PhysRevB.111.205117,li2025deeplearningshedslight,luo2025solvingfractionalelectronstates} can reach larger systems than ED, they might be helpful to characterize the quasiparticle/quasihole more precisely, clarify their response to band mixing in moir\'e materials, and even study non-Abelian excitations. Second, we only considered the short-range delta impurity potential in this work. It would also be useful to study more realistic long-range impurity potentials, such as the Gaussian potential and the Coulomb potential of a point charge above the sample~\cite{Johri-PhysRevB.89.115124}. Finally, we focused on the anyons of stable $\nu=2/3$ Laughlin FCIs. As the non-uniform quantum geometry is a distinctive feature of moir\'e bands compared to Landau levels, it is interesting to systematically study the quantum geometric effect on the interaction between mobile FCI anyons and the spatial structure of localized FCI anyons. For the latter, a density wave halo around anyons is predicted to appear when FCIs compete with charge density waves~\cite{PhysRevB.110.085120}.

\begin{acknowledgments}
We thank Nicolas Regnault and Bo Yang for helpful discussions. This work was supported by the National Natural Science Foundation of China (Grant No.~12374149,  12350403, and 12274333). F. W. is also supported National Key Research and Development Program of China (Grants No. 2021YFA1401300 and No. 2022YFA1402400). Part of the numerical calculations in this paper have been done on the supercomputing system in the Supercomputing Center of Wuhan University.
\end{acknowledgments}

\appendix

\section{Tilted sample}
\label{app::tilted}
In the system sizes tractable by ED, we often use the tilted geometry to make the sample to be as isotropic as possible. This is necessary for developing localized quasiparticles. The periodic boundary conditions of tilted samples are determined by two vectors ${\bm T}_1$ and ${\bm T}_2$, which are in general the integer superpositions of both ${\bm a}_1$ and ${\bm a}_2$. If ${\bm T}_i$ is simply proportional to ${\bm a}_i$ in both directions, we return to regular samples. In Table~\ref{Table:tilted}, we summarize the sample geometries used in this work.

\begin{table}
    \begin{ruledtabular}
        \begin{tabular}{ccc}
        $N_s$ & ${\bm T}_1$ & ${\bm T}_2$ \\ \hline 
        14 & (4,1) & (2,4) \\ \hline
        16 & (4,0) & (0,4) \\ \hline
        19 & (5,2) & (3,5) \\ \hline
        28 & (6,2) & (1,5) \\ \hline
        29 & (6,1) & (1,5) \\ \hline
        31 & (6,1) & (5,6) \\ \hline
        32 & (6,2) & (2,6) \\ \hline
        33 & (6,1) & (3,6) \\ \hline
        34 & (6,1) & (2,6) \\ \hline
        35 & (6,1) & (1,6) \\ \hline
        36 & (6,0) & (0,6) \\ \hline
        37 & (7,3) & (-3,4)
        \end{tabular}
    \end{ruledtabular} 
    \caption{Sample geometries used in our numerical simulations. The vectors ${\bm T}_1$ and ${\bm T}_2$ give the periodic boundaries of the sample. The two integers $(m,n)$ in the second and the third columns mean ${\bm T}_i=m{\bm a}_1+n{\bm a}_2$.}
    \label{Table:tilted}
\end{table}

\begin{figure}
\centerline{\includegraphics[width=\linewidth]{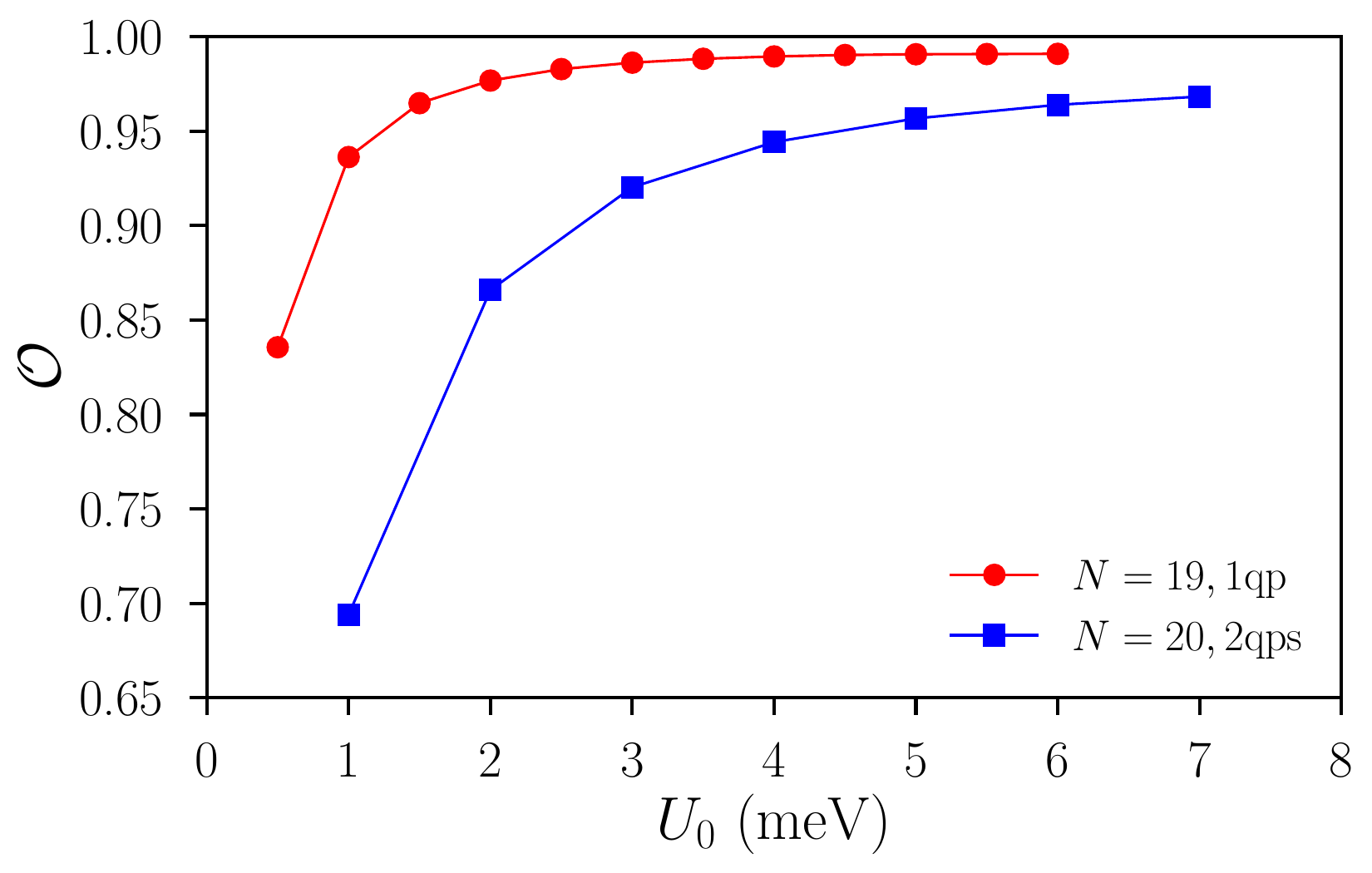}}
\caption{The overlap between the ground states obtained by diagonalizing the full tMoTe$_2$ Hamiltonian and those obtained by diagonalizing the impurity potential in the quasiparticle subspace. We consider two system sizes: $N=19, N_s=28$ and $N=20, N_s=29$, corresponding to the presence of one and two localized quasiparticles, respectively. The single quasiparticle is pinned at the sample center. The two quasiparticles are located at the sample center and ${\bm r}=(0,0)$. Single-band projection to the top valence band of $K$ valley is used.}
\label{overlap}
\end{figure}

\section{Diagonalize the full Hamiltonian of tMoTe$_2$}
\label{app::fullED}
In Sec.~\ref{sec:fci} of the main text, we first target on translational invariant tMoTe$_2$ systems with extra holes and moir\'e unit cells to get the subspace of delocalized quasiparticles, then diagonalize the impurity potentials to get the states with localized quasiparticles. Now we use an alternative method: we directly diagonalize the full Hamiltonian, including the band dispersion, interactions, and impurity potentials. Single-band projection in the $K$ valley is still imposed, and model parameters are the same as those in Sec.~\ref{sec:fci}. As the momentum is no longer a good quantum number when diagonalizing the full Hamiltonian, our calculations are limited to smaller system sizes than in Sec.~\ref{sec:fci}. 

In Fig.~\ref{overlap}, we compare the three ground states obtained by this alternative method with those in Sec.~\ref{sec:fci} for one and two localized quasiparticles. With the increasing of impurity potential strength $U_0$, the overlap grows quickly. For the single-quasiparticle case, the overlap has reached $\approx94\%$ when $U_0=1\ {\rm meV}$. However, $U_0=4\ {\rm meV}$ is needed for the two-quasiparticle case if we want to reach the same high overlap, probably because the two quasiparticles are not well separated from each other in the finite system. 

We hence estimate the impurity potential strength required to pin an isolate $\nu_h=2/3$ FCI quasiparticle as $\sim 1\ {\rm meV}$. This strength is still much smaller than the band gap of the top valence band in the single valley, which is $\approx 8.6 \ {\rm meV}$. In particular, we also notice that the overlaps are still growing even when $U_0$ has exceeded the gap separating the quasiparticle subspace and higher excited levels (Fig.~\ref{qp_energy}), indicating that the quasiparticle states cannot be mixed with other states by strong delta impurities. This also happens for the $\nu=1/3$ model Laughlin quasiholes in Eq.~(\ref{eq:Laughlinqh}), which remain the zero-energy ground states of the first-order Haldane pseudopotential combined with the delta pinning potential irrespective of the potential strength. Therefore, we expect that the quasiparticle state in tMoTe$_2$ is indeed close to the model Laughlin quasiparticle subject to density modulation, as shown in Eq.~(\ref{PhiLLh}). 

\begin{figure}
\centerline{\includegraphics[width=\linewidth]{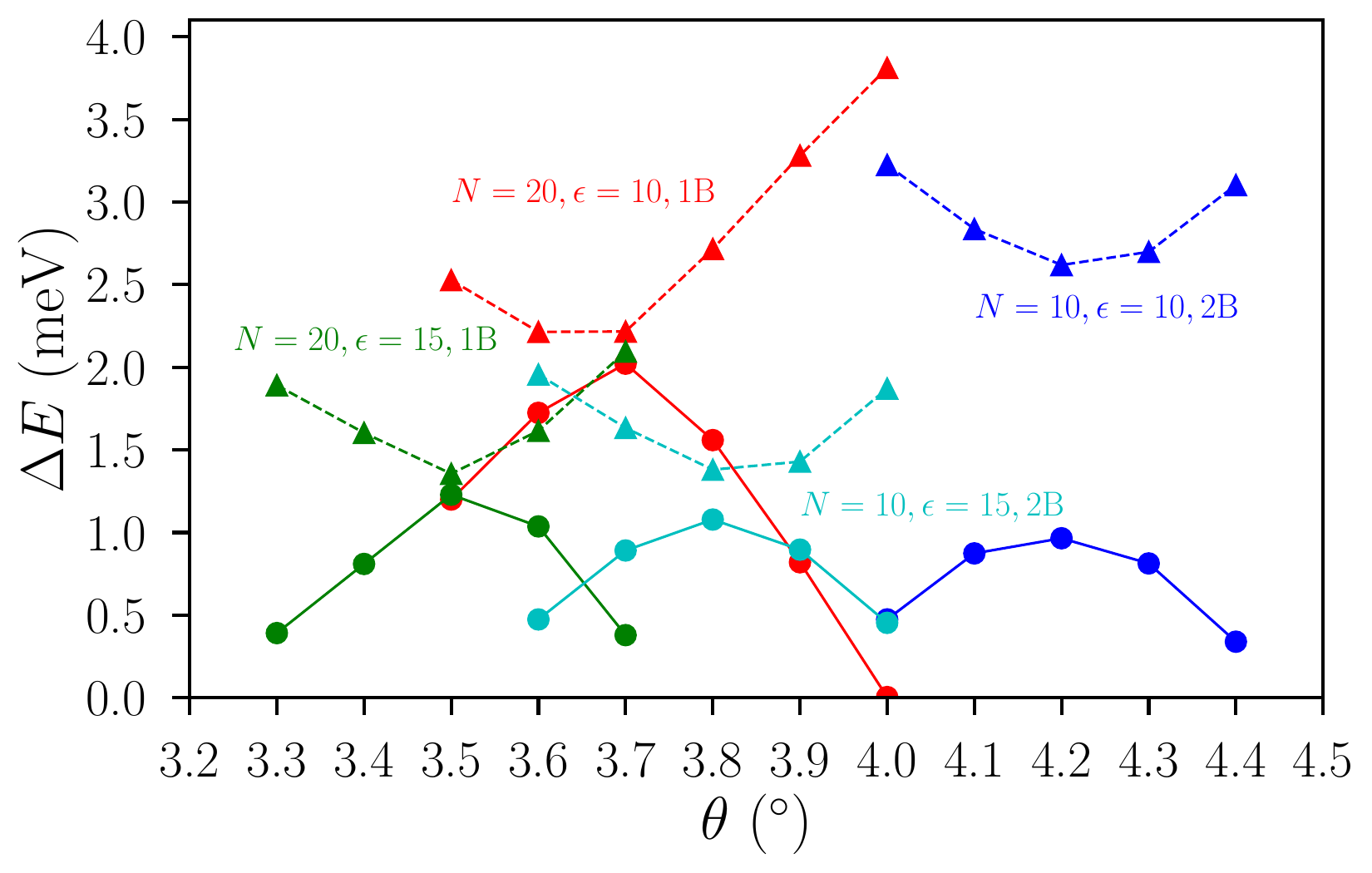}}
\caption{The energy gap (solid lines) and the splitting (dotted lines) of the two-quasiparticle manifold in tMoTe$_2$ as a function of the twist angle. ``1B'' means the projection to the top valence band in the $K$ valley, and ``2B'' means the projection to the top and the second valence bands in the $K$ valley. The system size and the dielectric constant are given in the texts near the corresponding data points. We do not restrict hole's occupation in the second valence band here.}
\label{qp_energy_more}
\end{figure}

\section{Dependence on dielectric constant}
\label{app::epsilon}

In Sec.~\ref{sec:fci} of the main text, we fix the dielectric constant as $\epsilon=10$. We have tried another value $\epsilon=15$. In Fig.~\ref{qp_energy_more}, we repeat the calculations of Fig.~\ref{qp_energy}(b) with this new parameter. Within single-band projection, we find the optimal twist angle, corresponding to the narrowest subspace of two delocalized quasiparticles, moves to $\theta=3.5^\circ$. The real-space density distribution around a single localized quasiparticle is very similar to that for $\theta=3.7^\circ, \epsilon=10$ (not shown here).

\begin{figure}
\centerline{\includegraphics[width=\linewidth]{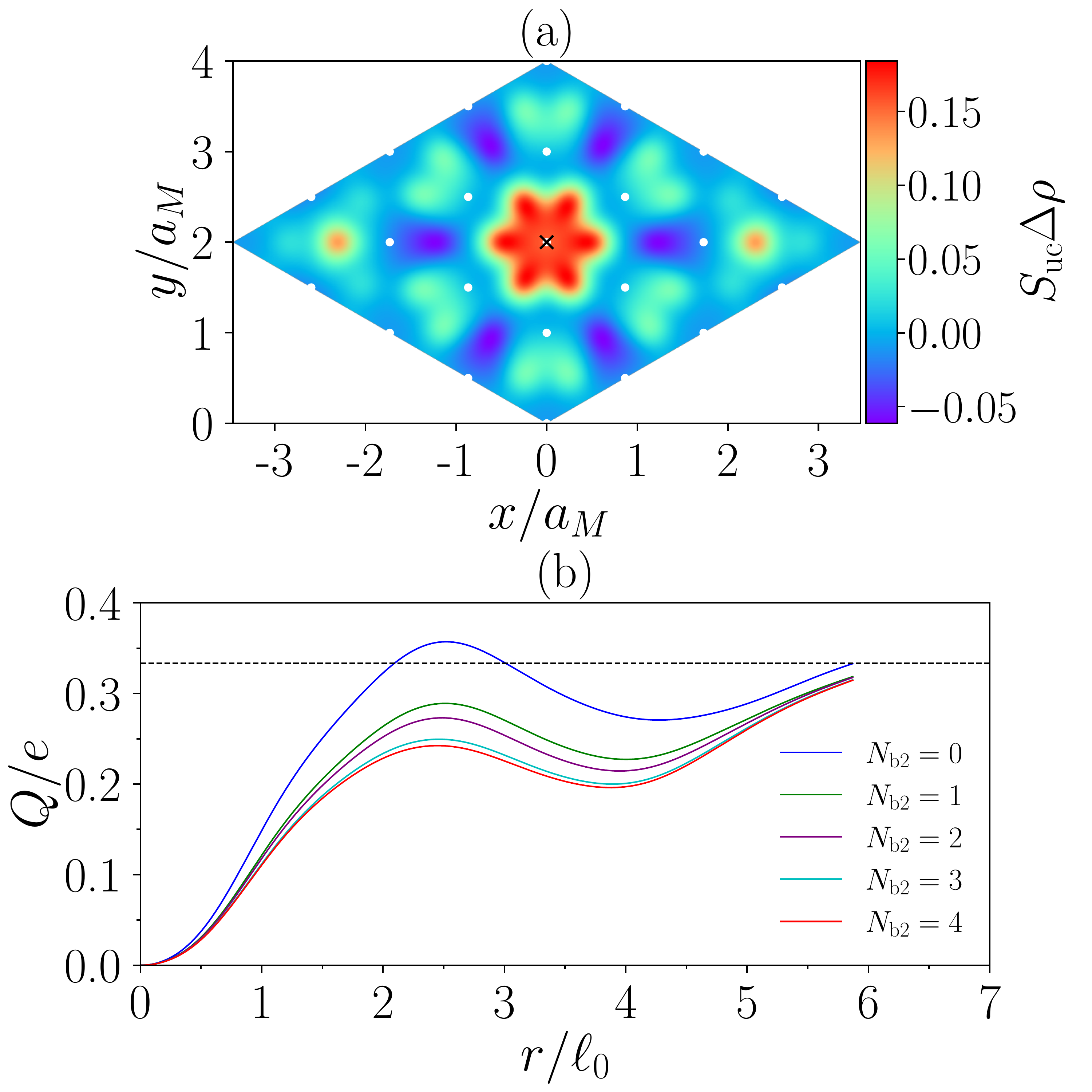}}
\caption{(a) Density profile around a single localized quasiparticle excitation in the $\nu_h=2/3$ tMoTe$_2$ FCI with $N=11, N_s=16$. We set $N_{\rm b2}=N$, so hole's occupation in the second valence band is unrestricted. The impurity potential (indicated by the black cross) is located at $R_M^M$. (b) Excess charge in the presence of a single localized quasiparticle excitation in the $\nu_h=2/3$ tMoTe$_2$ FCI with $N=13, N_s=19$. Here we compare the data obtained with different $N_{\rm b2}$. The horizontal line indicates $Q = e/3$, which is expected for a Laughlin-type quasiparticle.}
\label{FCI_1qp_more}
\end{figure}

\section{Two-band simulations}
\label{app::2B}

In Sec.~\ref{sec:fci} of the main text, we project the Hamiltonian to the $K$-valley top valence band only. For the model parameters chosen by us, we have checked that the delocalized quasiparticle subspace in the presence of either a single or two quasiparticles is still valley polarized if we keep the top valence bands in both valleys. 

In this section, to further explore the multi-band effects on FCI quasiparticles, we stick with the valley polarization, but keep the top and the second valence bands in the $K$ valley. The Hilbert space significantly grows when two bands are kept, so we can only reach much smaller systems compared with the single-band case. However, we still make efforts to study samples as large as possible by employing the ``band maximum'' technique introduced in Ref.~\cite{yu2024moirefractionalcherninsulators}. This method limits the number of holes occupying the second valence band while leaving the occupation in the top valence band unrestricted. We will gradually increase $N_{\rm b2}$, the number of holes in the second valence band, to include the band mixing in a controlled way. The single-band projection is recovered when $N_{\rm b2}=0$. For small $N_{\rm b2}$, the Hilbert space dimension is reduced a lot compared with the case of unrestricted hole occupation in the second valence band, so that the corresponding system sizes may become tractable for ED. Further reduction of the Hilbert space dimension can be realized by forbidding hole's occupation in specific momentum points, but we do not implement that in our work. 

Like in the main text, we will first obtain the quasiparticle subspace under two-band projection with a specific $N_{\rm b2}$, then diagonalize the impurity potential within this subspace to localize the quasiparticles. In Fig.~\ref{qp_energy_more}, we display the energy gap and splitting of the quasiparticle subspace in the presence of two delocalized quasiparticles for $N=10, N_s=14$. In this case we allow all holes occupying the second valence band, i.e., $N_{\rm b2}=N$. The optimal twist angle corresponding to the smallest spreading is $4.2^\circ$ and $3.8^\circ$ for $\epsilon=10$ and $\epsilon=15$, respectively. In both cases we obtain similar results. We will focus on $\epsilon=15,\theta=3.8^\circ$ in what follows.

In Fig.~\ref{FCI_1qp_more}(a), we display the profile of a single quasiparticle for $N=11, N_s=16, N_{\rm b2}=N$, with the impurity potential located at the $R_M^M$ point. This is the largest system we can reach under two-band projection if the hole's occupation in the second valence band is unrestricted. The density has been averaged over the three nearly degenerate states with a single localized quasiparticle. Clearly the quasiparticle is not well developed for this system size. Compared with Fig.~\ref{FCI_1qp}, the density discrepancy still strongly oscillates in the entire sample. In Fig.~\ref{FCI_1qp_more}(b), we gradually increase $N_{\rm b2}$ in a larger system size $N=11, N_s=19$ to demonstrate how the localized quasiparticle is affected by the band mixing. In the limit of $N_{\rm b2}=0$, the excess charge as a function of distance $r$ from the impurity potential is very similar to that shown in Fig.~\ref{FCI_1qp}(d). With the increasing of $N_{\rm b2}$, the first peak of $Q(r)$ is gradually suppressed, such that $Q(r)$ does not oscillate around $e/3$ at the available values of $r$. We notice that the curve does not change much from $N_{\rm b2}=3$ to $N_{\rm b2}=4$, so probably $N_{\rm b2}=4$ is already close to the case of $N_{\rm b2}=N$. 

\begin{figure}
\centerline{\includegraphics[width=\linewidth]{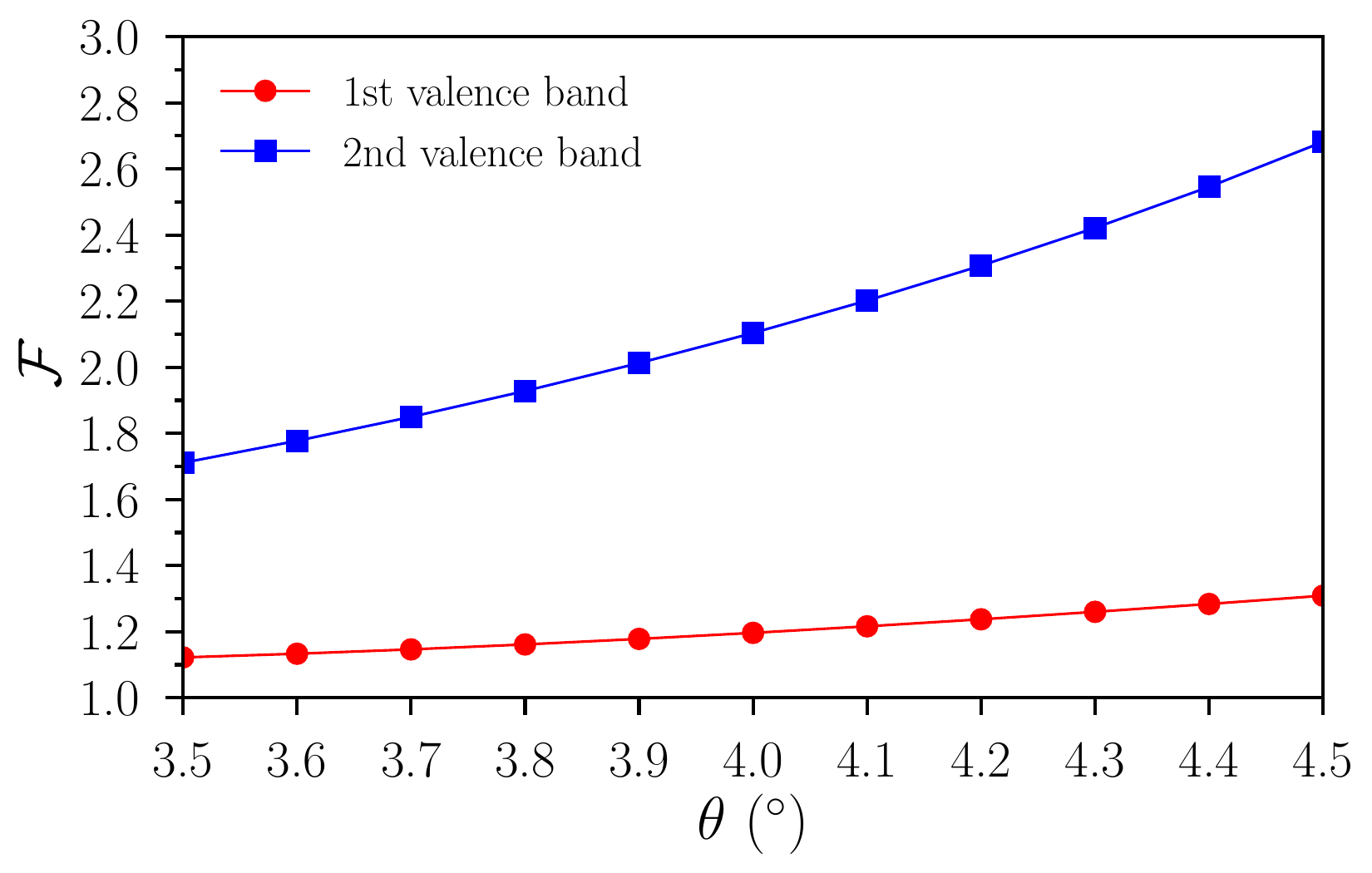}}
\caption{The quantity $\mathcal{F}$ defined in Eq.~(\ref{eq::trg}) as a function of twist angle $\theta$ for the top and the second valence bands of tMoTe$_2$ in the $K$ valley.}
\label{trg}
\end{figure}

The appreciable suppression of the first peak of $Q(r)$ was also observed using the delta impurity potential for the $\nu=1/3$ Laughlin state of electrons in the second Landau level~\cite{PhysRevLett.110.186801,Johri-PhysRevB.89.115124}. So it may be useful to compare the top and the second tMoTe$_2$ valence bands with conventional Landau levels. In the aspect of single-particle quantum geometry, a convenient indicator of their similarity is 
\begin{eqnarray}
\mathcal{F}\equiv\min_{\tilde g}\left(\frac{1}{2\pi}\int_{\rm MBZ}{\rm tr}_{\tilde g} g({\bm k})\right),
\label{eq::trg}
\end{eqnarray} 
where $g({\bm k})$ is the quantum metric of a specific band. For the LLL and the second Landau level, $\mathcal{F}$ takes the value of $1$ and $3$, respectively. In Fig.~\ref{trg}, we show $\mathcal{F}$ for the top and the second valence bands of tMoTe$_2$ as a function of $\theta\in[3.0^\circ,4.5^\circ]$. In the entire range of $\theta$, we find for both bands that the minimum in Eq.~(\ref{eq::trg}) is always reached when ${\tilde g}$ is a $2\times 2$ identity matrix. For the top valence band, $\mathcal{F}$ only grows from $1.1$ to $1.3$ with increasing $\theta$, suggesting the geometric similarity with the LLL. By contrast, $\mathcal{F}$ of the second valence band grows more quickly from $1.7$ to $2.7$ with increasing $\theta$, which implies that the second valence band might be a superposition of the lowest and the second LLs~\cite{liu2024theorygeneralizedlandaulevels}. Therefore, the band mixing in tMoTe$_2$ could be analogous to the mixing of the LLL and the second Landau level.

It requires more advanced numerical simulations in sufficiently large systems to conclude whether the suppression of the first peak of $Q(r)$ in the presence of band mixing is a finite-size effect, or a signal of larger spatial extension of the quasiparticle. Ref.~\cite{Johri-PhysRevB.89.115124} found that long-range impurity potentials (like the Gaussian potential, or the Coulomb potential of a point charge above the sample) perform better than the delta potential to localize a $\nu=1/3$ Laughlin quasihole in the second Landau level, but even in that case the obtained spatial extension of the quasihole is $1.75$ times larger than that in the LLL. Since the second valence band of tMoTe$_2$ may have significant weight on the second Landau level, the band mixing in tMoTe$_2$ could also increase the size of the $\nu_h=2/3$ FCI quasiparticle. 

\section{Localized $\nu=2/3$ quasiholes}
\label{app::qh}

\begin{figure}
\centerline{\includegraphics[width=\linewidth]{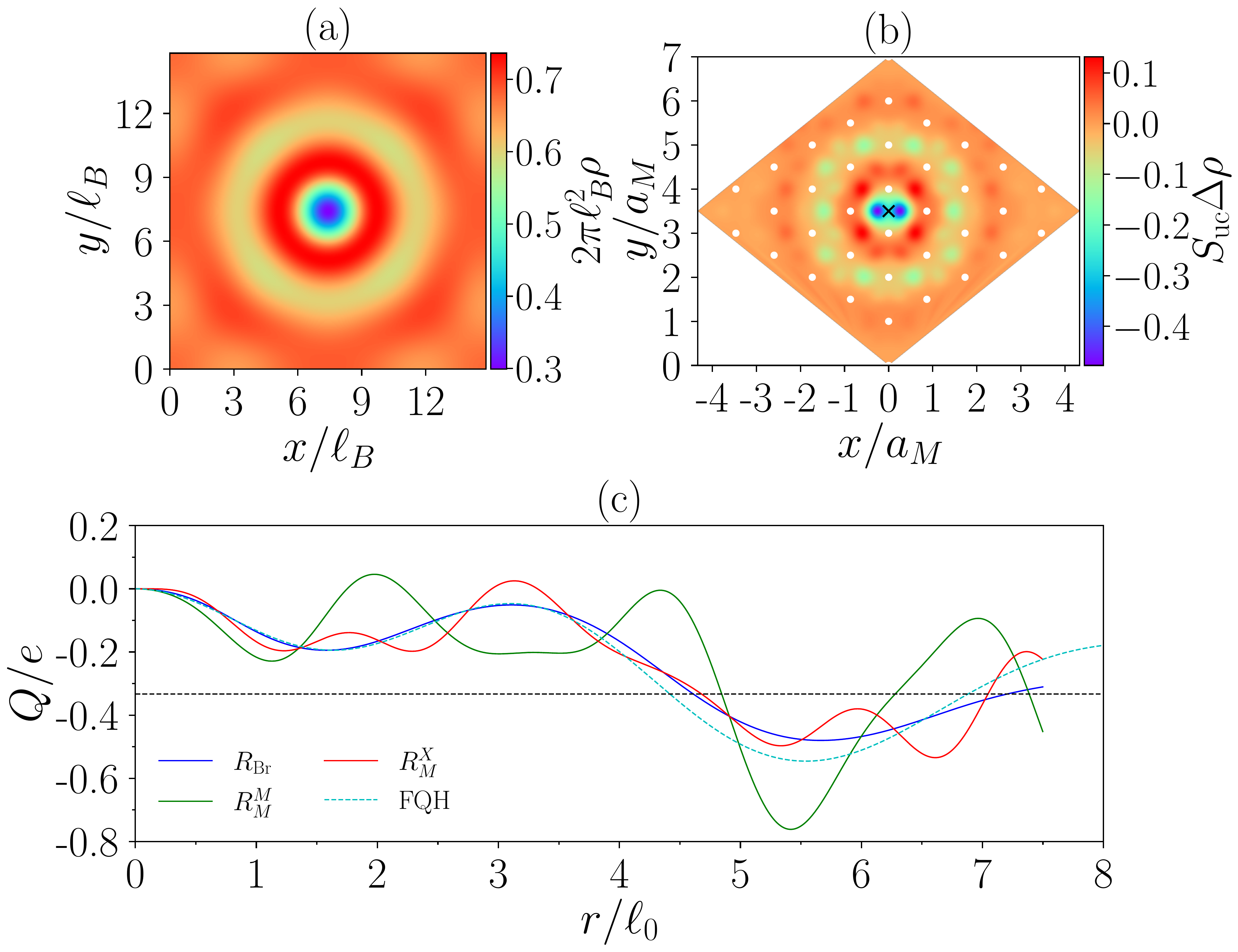}}
\caption{(a), (b) The density distribution for a localized single-quasihole excitation created in (a) the $\nu=2/3$ FQH state on a square torus with $N=23,N_\phi=35$ and (b) the $\nu_h=2/3$ FCI in tMoTe$_2$ with $N=23,N_s=35$. In both cases the impurity potential is located at the center of the sample, which coincides with the $\mathcal{R}_{\text{Br}}$ position in tMoTe$_2$. (c) Excess charges around the impurity potential. In the FCI case we consider different positions of the impurity potential, as indicated in the legend. The horizontal reference line indicates $Q=-e/3$, which is expected for a Laughlin-type quasihole. We use the screened Coulomb interaction for both FQH and FCI systems.}
\label{qh}
\end{figure}

In the main text, we have characterized localized $\nu=2/3$ quasiparticles for both the FQH and FCI cases. For completeness, we now present the results of a single localized $\nu=2/3$ quasihole. The calculations are similar to those in Secs.~\ref{continuuma} and \ref{sec:fci_1qp}, with the same model details. We use the LLL projection and the single-band projection for the FQH and FCI system, respectively. The subspace of a single mobile $\nu=2/3$ quasihole is created when the system size satisfies $N_{\phi/s}=\frac{1}{2}(3N+1)$. We then diagonalize a repulsive delta impurity potential to pin the quasihole. Three nearly degenerate ground states are obtained.

The density profiles in the LLL and tMoTe$_2$ are displayed in Figs.~\ref{qh}(a) and \ref{qh}(b), respectively, which have been averaged over the three nearly degenerate ground states. In both cases we indeed observe density minima near the impurity potential, which is expected for a localized quasihole. However, as shown in Figs.~\ref{qh}(c), the excess charge in the quasihole case is much messier than that in the quasiparticle case [Fig.~\ref{FCI_1qp}(d)]. In particular, the excess charge does not start regular oscillations around $-e/3$ at relatively small $r$. It is first stuck between about $-0.2e$ and $0.05e$ until $r\sim 4\ell_{B/0}$. Even after that we cannot see a clear tendency of convergence to $-e/3$. Moreover, $Q(r)$ in tMoTe$_2$ now shows strong dependence on the position of the impurity potential. The data indicate that the spatial extent of a localized $\nu=2/3$ quasihole is larger than that of a localized $\nu=2/3$ quasiparticle, so that we cannot characterize it well in finite systems reached by ED. In the context of FQH states, it has been known that the localized $\nu=1/3$ quasiparticle (equivalently, the $\nu=2/3$ quasihole) has a bigger size because of its complicated internal structure~\cite{KJONSBERG1999705,PhysRevLett.112.026804}.

\bibliography{FCI_hole}

\end{document}